\documentclass{str-author}

\usepackage{subcaption}

\begin{document}
\runhead{A single camera 3D DIC system for the study of adiabatic shear bands}{T. G. White et al.}

\title{A single camera three-dimensional digital image correlation system for the study of adiabatic shear bands}

\author{T. G. White, J. R. W. Patten, K. Wan, A. D. Pullen, D. J. Chapman and D. E. Eakins}

\address{Institute of Shock Physics, Imperial College London, London, UK} 

\abstract{
We describe the capability of a high resolution three-dimensional digital image correlation (DIC) system specifically designed for high strain-rate experiments. Utilising open-source camera calibration and two-dimensional digital image correlation tools within the MATLAB framework, a single camera 3-D DIC system with sub-micron displacement resolution is demonstrated. The system has a displacement accuracy of up to 200 times the optical spatial resolution, matching that achievable with commercial systems. The surface strain calculations are benchmarked against commercially available software before being deployed on quasi-static tests showcasing the ability to detect both in- and out-of-plane motion. Finally, a high strain-rate (1.2$\times$10$^3$~s$^{-1}$) test was performed on a top-hat sample compressed in a split-Hopkinson pressure bar in order to highlight the inherent camera synchronisation and ability to resolve the adiabatic shear band phenomenon. 
}


\maketitle

\section{Introduction}
\label{sec:intro}

Full-field deformation measurements are an important component in the analysis of material behaviour when subjected to both quasi-static and high-rate loading.  Unlike traditional methods such as load cells, which provide global information, or strain gauges, which yield information averaged over a limited portion of the sample, optical techniques are a non-contact method able to provide point-wise information over a large area of the sample. This becomes particularly important when investigating inhomogeneous behaviour or systems in hostile environments that necessitate a non-contact measurement. 

Developed in the 1980s \cite{chu,sutton}, Digital Image Correlation (DIC) is a non-contact optical method capable of measuring the full-field strain distribution over a sample surface. The relatively lenient requirements necessary to perform DIC has lead to widespread use and popularity beyond competing optical techniques such as Moir\'{e} interferometry \cite{interferometry} or Holography \cite{holography}. Two-dimensional (2-D) DIC utilises a single camera to measure in-plane sample deformation through cross-correlation of sequential greyscale images and is capable of obtaining displacement vectors with up to 0.01 pixel precision \cite{suttonbook}. Since its conception DIC has become an essential technique in experimental/solid mechanics as well as having uses in material characterization, architecture, aerospace and biology \cite{suttonbook}. 

\begin{figure*}
\center
\resizebox{0.9\textwidth}{!}{
 \includegraphics{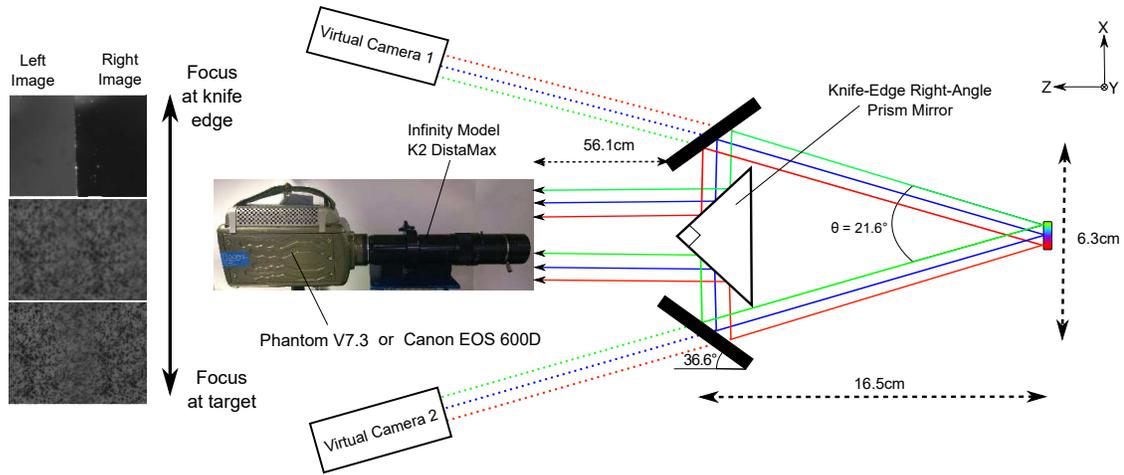} }
\caption{Schematic of experimental layout showing approximate distances between the target, mirror system and high speed camera. The inset on the left shows the blurring of the knife-edge mirror as focus is shifted from the knife-edge to the target.}
\label{fig:exp}       % Give a unique label
\end{figure*}

Two-dimensional DIC is limited to measuring the in-plane deformation of planar samples, hence requiring the camera to be located normal to the target surface. However, sample deformation is rarely limited to in-plane motion and more general target geometries require three-dimensional (3-D) position and displacement measurements. In this work this is achieved through 3-D DIC, that is stereoscopic imaging of the sample using two calibrated cameras. Combining knowledge of the relative camera angles with standard 2-D DIC reveals the out-of-plane deformation \cite{luo}. 

Many commercial systems for 3-D DIC now exist (VIC-3D \cite{vic3d}, ARAMIS \cite{aramis}, DaVIS \cite{davis}), however, both the hardware and software can be typically prohibitively expensive for smaller research groups. In addition, these closed source codes prevent modification or improvement by the user and can lead to uncertainty over the precise details of the algorithms implemented. 

Performing 3-D DIC at high strain-rate introduces further complications and expense as this typically necessitates the use of two synchronized high speed cameras. In such systems the two greatest costs are the high frame-rate cameras and the bespoke software required for stereo camera calibration and digital image correlation, while the electronic synchronization of the two cameras provides the largest source of temporal error. To this end the development of a single camera 3D-DIC system, which eliminates issues surrounding high rate camera synchronization and offers improved stability, has received considerable attention. However, these attempts place stringent requirements on target geometry \cite{singlecamera2,singlecamera3,singlecamera4}, require precise camera and optical set-ups \cite{Prentice} or are analyzed with commercial software \cite{singlecamera1}. Here we aim to demonstrate a low-cost 3-D DIC system capable of achieving high rate, high spatial resolution, three-dimensional deformation measurements.

In the following study we develop a single camera 3-D DIC system which alleviates synchronization errors and is based on freely available open source codes. The camera calibration was carried out using the Caltech Calibration Toolbox for MATLAB \cite{CameraCal} while the 2-D DIC is achieved using Ncorr developed at Georgia Institute of Technology \cite{Ncorr1, Ncorr3}. The Ncorr system has previously demonstrated excellent agreement with commercial software \cite{Ncorr2}. 

The aim of this work is to assess the applicability of using a single camera together with open-source software to achieve three-dimensional high-rate, high spatial and displacement resolution DIC, taking advantage of the improved temporal synchronisation and stability associated with single camera systems, the flexibility and transparency associated with using open-source codes and at a fraction of the cost of a dual-camera commercial system. 

This system has been tailored for a field of view of a few millimeters as this allows the study of adiabatic shear bands (ASBs), a widely encountered phenomenon in engineering alloys subjected to high strain-rate loading \cite{bdodd}. They form on the sub-millimeter scale and are of particular interest to the aerospace industry where shear bands often occur when a hard object impacts the rotating fan blades at high speed; the velocity of impact leads to low thermal conduction and an increase in temperature, thermal softening and ultimately in catastrophic failure from brittle-like fracture  \cite{aerospace}. The study of ASBs is one example of a situation that requires high-rate 3-D DIC, however the approach discussed here is broadly applicable, and can be tailored to different applications by simple exchange of camera lens and optical components. 

The need for 3-D DIC in the study of ASBs is not immediately obvious as they are typically a 2D phenomenon when viewed on the surface of the sample. However, with a 2-D system the camera must be placed orthogonal to the sample surface which is assumed to move only in-plane, this constraint is relaxed in the 3-D setup as the surface normal is explicitly measured. A 3-D system thus allows measurements to be taken despite sample motion (e.g. tilt, bulging, necking) and when non-planar sample geometries are used. Furthermore, ASBs are by definition a localization phenomenon and therefore marks the material deformation in a non-homogeneous manner with regions of large plastic deformation \cite{bdodd}. Measurement of the out-of-plane displacement allows the full finite-strain Lagrange strain tensor to be calculated \cite{mechconmed}. Retaining all terms in the Lagrange strain tensor ensures the large finite strains present in ASBs are appropriately described.

\section{Experimental Set-up}
\label{sec:exp}

\begin{table*}
\center
\caption{Properties of the camera set-ups used in the displacement precision calibration, see Figures \ref{fig:displc} and \ref{fig:displp}. Each case utilised the K2 DistaMax lens (standard configuration), however the Phantom V7.3 with reduced resolution was used with an additional $4\times$  magnification to compensate for the reduction in the field of view. The optical resolution was obtained through imaging a knife-edge object and using the 10\% - 90\% intensities of the edge spread function \cite{stevesmith}.}
\label{tab:1}       % Give a unique label
\begin{tabular}{lllll}
\hline\noalign{\smallskip}
Camera &  CCD Size (pixels) & Pixels per mm & Field of View (mm) & Optical Resolution ($\mu$m) \\
\noalign{\smallskip}\hline\noalign{\smallskip}
Canon EOS 600D & 5184$\times$3456 & 620 &  4.1$\times$5.5 & 97\\
Phantom V7.3 & 800$\times$600 & 110 &  3.6$\times$5.4 & 155\\
Phantom V7.3 & 240$\times$184 & 30 &    3.6$\times$5.4 & 300\\
\noalign{\smallskip}\hline
\end{tabular}
\end{table*}

The general approach uses reflective optics to image a sample surface from two viewing-angles onto a single camera. The set-up consists of two planar 25~mm square mirrors along with a single knife-edge right-angled prism mirror; these were used to create two views of the target on a single CCD, see Figure \ref{fig:exp}. Both the prism mirror and side mirrors are separately mounted such that they can be moved independently to adjust the image on the camera. As with most commercial set-ups the system is relatively insensitive to the exact angles used due to the robustness of the camera calibration method, the angles do not need to be known accurately \emph{a-priori}. However, the angle subtended by the two mirrors and the sample (marked $\theta$ in Figure \ref{fig:exp}) should remain between $20^\circ-30^\circ$ to achieve a good balance between out-of-plane precision and image correlation. A larger angle allows the out-of-plane motion to be determined with greater accuracy while conversely making the image correlation between the two images more difficult due to increased image disparity \cite{singlecamera1}. The distances and angles used in this work are given in Figure \ref{fig:exp}, as are images showing a typical field of view as focus is shifted from the knife-edge to the target. Future work could involve extending the ideas presented here to more than two images, and hence viewing angles, helping to constrain the problem further and improve accuracy.

In order to demonstrate the ability to measure full-field displacement of both in- and out-of-plane motion from a single camera the system was set-up to observe a flat speckled surface. The speckle pattern was applied using an Iwata HP-B Plus airbrush with a 200~$\mu$m nozzle using Daler Rowney FW Acrylic Artist's Ink, Black 028. This ink was found to adhere best to the sample surface. The pressure was varied to alter the size of the droplets; a pressure of 1~bar was found to give an optimal speckle. In this case an optimal speckle is defined as being without large featureless areas which reduce the achievable spatial resolution and without fine marks below the optical resolution of the set-up. A nozzle to sample distance of 23~cm was used. A good review of optimal speckle patterns for DIC can be found in reference \cite{speckle}. A typical speckle pattern can be seen in the insert of Figure \ref{fig:exp}. 

This target was mounted on a Thorlabs PT3-Z8 translation stage providing XYZ motion while the motorised Z825B actuators provide up to 25~mm travel with a minimum resolution of 29~nm. A camera (either a digital SLR or high-speed camera) was attached to an Infinity K2 DistaMax long-distance microscope lens (standard configuration, with $4\times$ magnification provided by two NTX tube amplifiers) to allow ultra-high resolution images to be taken. The entire experiment was performed on a vibrationally-damped optical table to isolate the set-up from any external vibrations. The high frame-rate Phantom is capable of taking 6600 frames a second utilising all 800x600 pixels and up to 50000 frames per second at a reduced pixel resolution of 240x184. Table \ref{tab:1} lists the parameters of the three different set-ups evaluated in this work. 

\section{Reconstructing the 3-D world}
\label{sectiontheory}

\begin{figure*}[t!]
\center
\resizebox{0.8\textwidth}{!}{
 \includegraphics{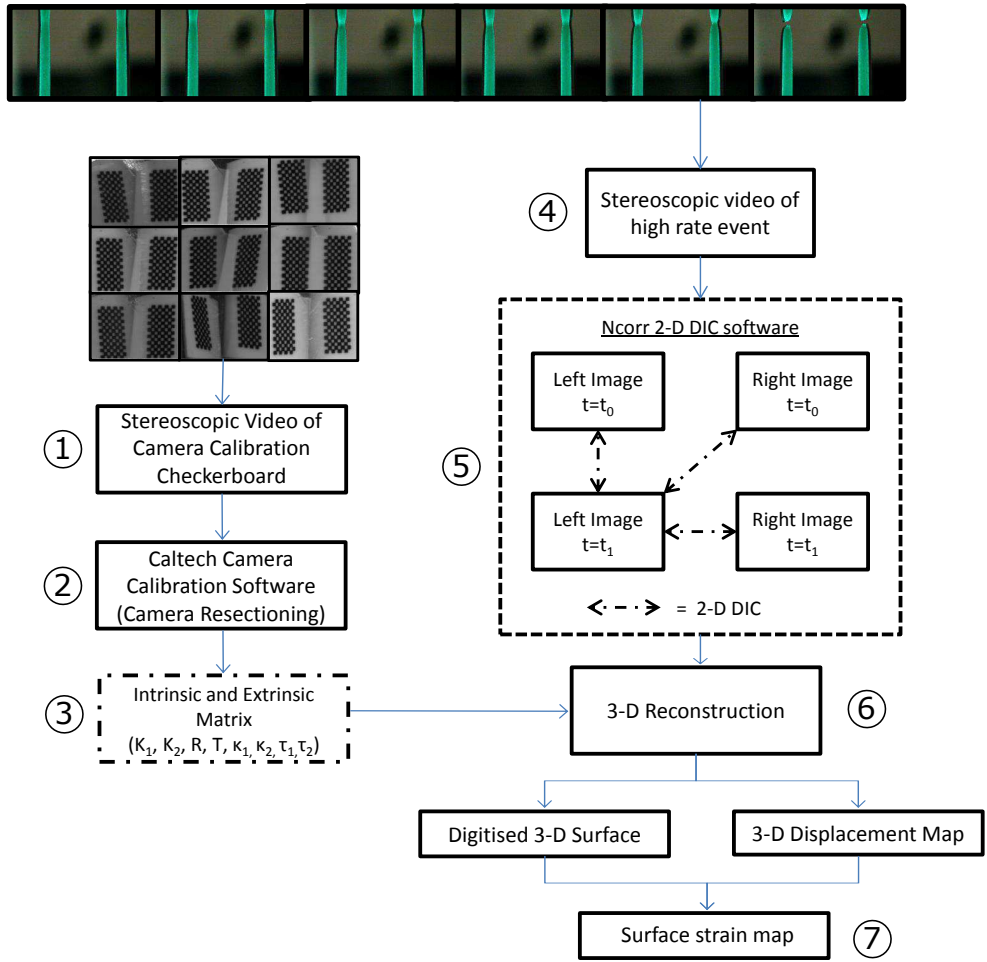} }
\caption{Workflow diagram showing the preprocessing of the calibration images by the CalTech Camera Calibration software \cite{CameraCal} and the Ncorr 2-D DIC software \cite{Ncorr1} before being processed by the 3-D reconstruction algorithm detailed in the text.}
\label{fig:workflow}       % Give a unique label
\end{figure*}

The majority of this work revolves around reconstruction of the 3-D world from a pair of stereoscopic images. The general method is outlined in Figure \ref{fig:workflow}, with steps 1-3 representing calibration of the system, steps 4-5 digital image correlation between pairs of images, step 6 combines these results with the calibration to produce a 3-D surface and displacement map which are then both used to produce a surface strain map. This work utilises the pinhole camera model for the projective mapping from 3-D world coordinates, denoted $\mathbf{P_w}=$[$X_w Y_w Z_w 1$],  to the two-dimensional camera coordinates, denoted $\mathbf{C_1}=$[$u_1 v_1 1$]. Mathematically this is expressed as the matrix multiplication,

\begin{equation}
\label{camera1}
\mathbf{C_1}=
\begin{bmatrix}  u_1 \\ v_1 \\ 1 \end{bmatrix} 
=\mathbf{K_1} \cdot 
\begin{bmatrix}  \mathbf{R_1} |\mathbf{ T_1}  \end{bmatrix} 
\cdot
\mathbf{P_w}\,,
\end{equation} where $K_1$ is the intrinsic matrix of the camera given by,

\begin{equation}
\mathbf{K_1}= \begin{bmatrix} f_x & s & x_0 \\ 0 & f_y & y_0 \\ 0 & 0 & 1\end{bmatrix}.
\end{equation} 

The intrinsic matrix contains the coordinates of the principal point of the camera, $(x_0, y_0)$, the horizontal and vertical focal lengths of the camera $(f_x. f_y)$ given in units of pixel dimensions, and $s=\alpha_c{f_x}$ which allows for non-rectangular pixels through the angle $\alpha_c$. The terms $\mathbf{R_1}$ and $\mathbf{T_1}$ together define the $3\times4$ extrinsic matrix of camera 1 and relates the position of the camera to a particular world coordinate frame, through a rotation followed by a rigid body translation. For a stereo system, i.e. a system where two cameras are used to resolve a single scene, an equivalent equation can be written for the second camera,

\begin{equation}
\label{camera2}
\mathbf{C_2}= \mathbf{K_2} \cdot
\begin{bmatrix}  \mathbf{R_2} |\mathbf{ T_2}  \end{bmatrix} 
\cdot
\mathbf{P_w}.
\end{equation} 

Throughout this work the reference frame of camera 1 is used as the world reference frame, hence, $\mathbf{R_1}=\mathbf{I}$, $\mathbf{T_1}=\mathbf{0}$, $\mathbf{R_2}=\mathbf{R}$ and $\mathbf{T_2}=\mathbf{T}$. Therefore equations \eqref{camera1} and \eqref{camera2} together represent a system of four simultaneous equations. Hence, it is possible with knowledge of the position of a single point on each camera, $u_1,v_1,u_2,v_2$ the equations can be solved to obtain the position of that point in the world reference frame, $\mathbf{P_w}$. The identification of equivalent points within each image is achieved through standard 2-D DIC.  

If the imaging system contains distortion then the reconstructed 3-D world, represented by the points $\mathbf{P_w}$, will also suffer from distortion. However, the camera calibration toolbox is naturally able to identify distortion in the imaging system through the chequerboard calibration method \cite{CameraCal}. Radial distortion is parameterized through the two parameters $\kappa_1$ and $\kappa_2$ while tangential distortion is included through $\tau_1$ and $\tau_2$. The distorted world coordinates, $\mathbf{P_w}$, can then be related to the undistorted coordinates, $\mathbf{P'_w}$, through Equations \eqref{distortion} and \eqref{tangdist}, where $x=X'_w/Z'_w$ and $y=Y'_w/Z'_w$ are the reduced undistorted coordinates, $r^2=x^2+y^2$ and $\mathbf{T_d}$ the translational component of distortion.
 
\begin{equation}
\label{distortion}
\begin{bmatrix} X_w/Z_w \\ Y_w/Z_w \end{bmatrix} = \begin{bmatrix} x \\ y \end{bmatrix} (1+\kappa_1{r^2}+\kappa_2{r^4}) + \mathbf{T_d} ,
\end{equation} 
 
\begin{equation}
\label{tangdist}
\mathbf{T_d}=\begin{bmatrix} 2\tau_1{xy} + \tau_2(r^2 2x^2) \\ 2\tau_2{xy} + \tau_1(r^2 2y^2) \end{bmatrix} .
\end{equation} 

Steps 1-3 represent the process of camera resectioning, that is the technique by which both the intrinsic ($\mathbf{K}$ - focal length, pixel properties and distortion) as well as the extrinsic ($\mathbf{T}$ - translational and $\mathbf{R}$ - rotational displacements) parameters of two or more cameras can be found. It is a crucial component of any stereoscopic imaging technique and this work utilises the Caltech Calibration Toolbox \cite{CameraCal} as well as the inbuilt Stereo Calibration function in MATLAB \cite{matlab}. Calibration in both methods is achieved using multiple images of a known, and well defined, two-dimensional pattern. Typically an asymmetrical chequerboard is used. Both methods used here utilise a non-linear minimisation on the re-projected chequerboard to find the unknown coefficients through a steepest gradient descent method \cite{CameraCal,matlab}. The multiple chequerboard images enable the 3-D space to be mapped onto the 2-D images. Typically, the re-projection errors, i.e. the offset between a chequerboard corner and the re-projection of the corner using the pin-hole camera formalism are often found to be sub-pixel in all but the most distorted systems. The chequerboard calibration technique contains a single free parameter, the chequerboard width, which sets the scale of the image.

\begin{figure}
        \centering
        \begin{subfigure}[b]{0.8\columnwidth}
                \includegraphics[width=\textwidth]{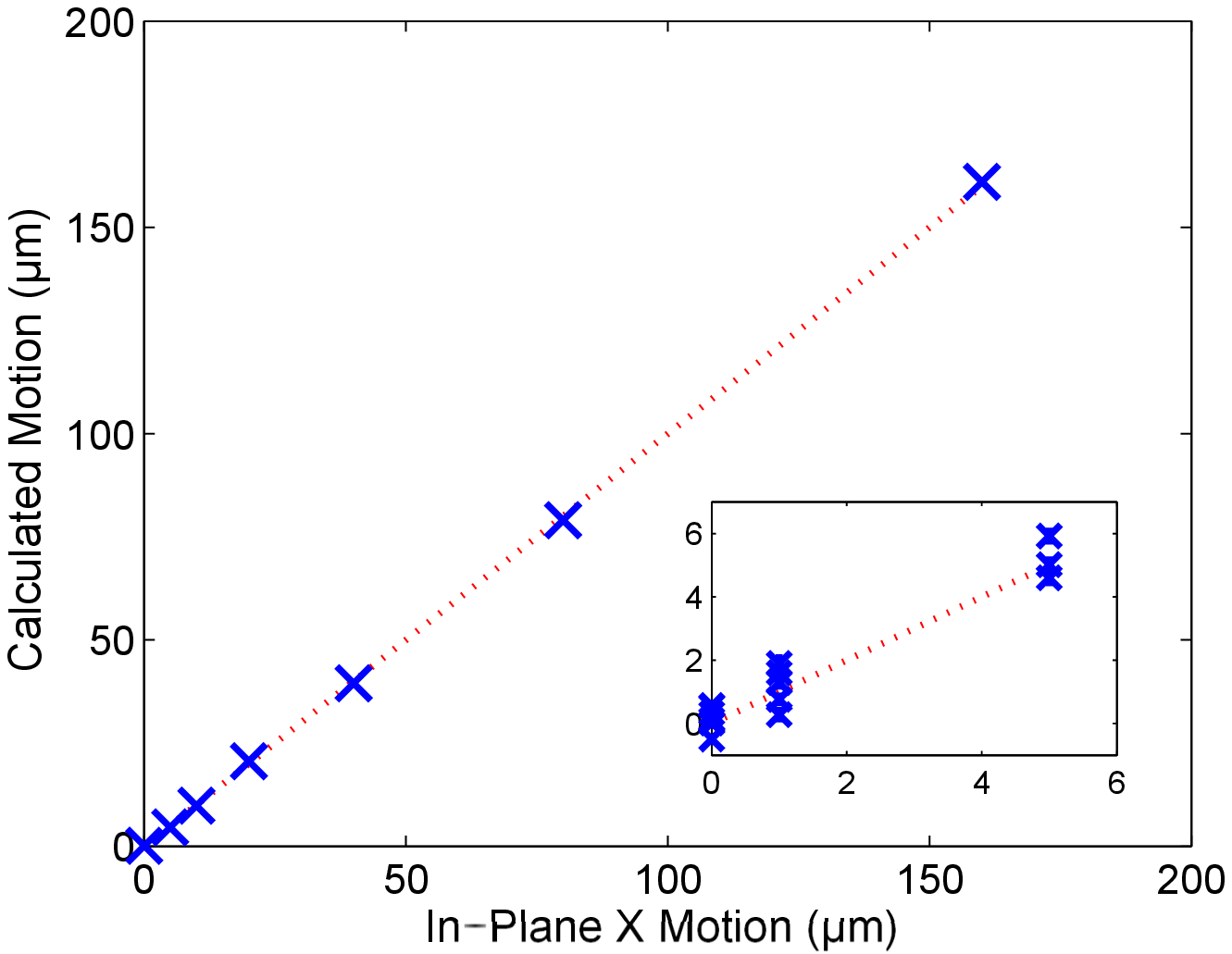}
                \caption{x-displacement}
        \end{subfigure}%
        ~ %add desired spacing between images, e. g. ~, \quad, \qquad, \hfill etc.
          
        \begin{subfigure}[b]{0.8\columnwidth}
                \includegraphics[width=\textwidth]{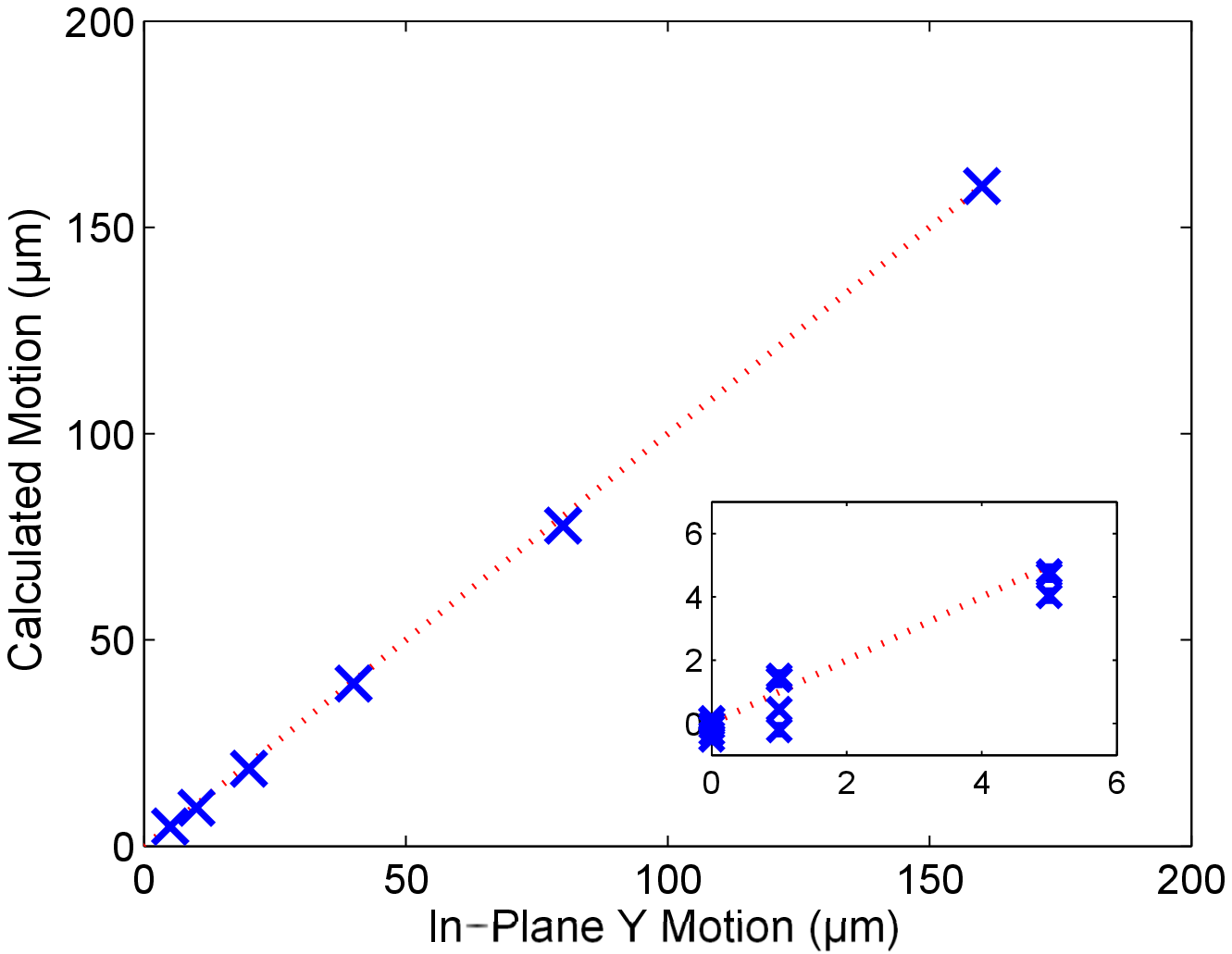}
                \caption{y-displacement}
        \end{subfigure}
        ~ %add desired spacing between images, e. g. ~, \quad, \qquad, \hfill etc.
          
        \begin{subfigure}[b]{0.8\columnwidth}
                \includegraphics[width=\textwidth]{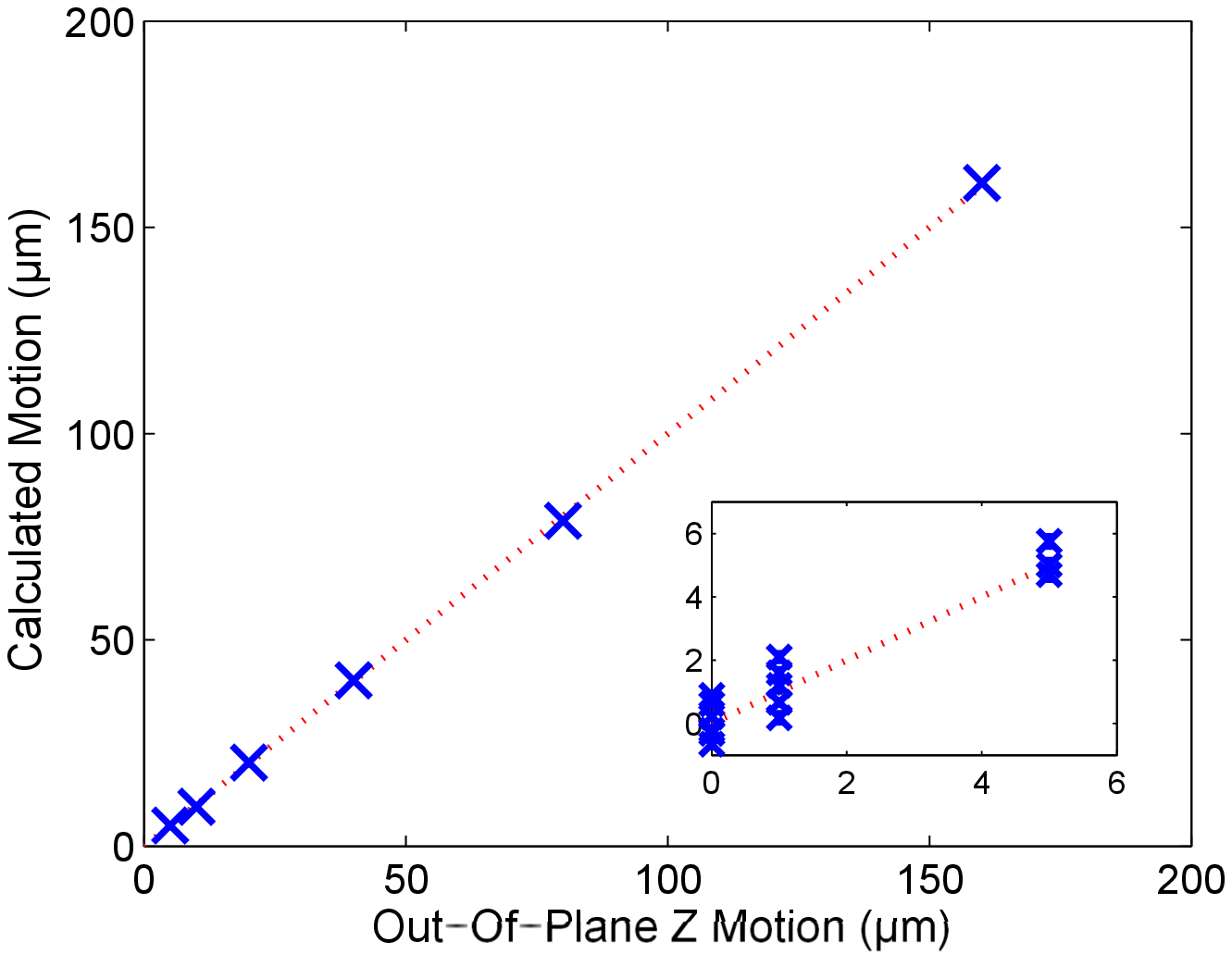}
                \caption{z-displacement}
        \end{subfigure}
        \caption{Displacement accuracy achieved with Canon EOS 600D with 5184$\times$3456 pixels. The results plotted represent an average over the spatial extent of the sample with error bars showing the standard deviation. The error bars are not visible on this scale. The dotted line shows $y=x$. The inset shows a close up of the 1 and 5 micron displacements where the error and fluctuations are clearer.  }\label{fig:displc}
\end{figure}

Steps 4-5 of the 3-D reconstruction is based on identifying the pixel position ($\mathbf{C}$ - $u_1$,$v_1$,$u_2$,$v_2$) of equivalent points in the set of two stereoscopic images, this is achieved through standard 2-D DIC techniques. Ncorr is an open-source MATLAB program for 2-D DIC which runs entirely inside the MATLAB environment; this enables easy compatibility with the camera calibration methods discussed above, but additionally utilises C++ and MEX to improve efficiency \cite{Ncorr1}. Ncorr uses the inverse compositional method which is fast, robust and accurate compared to more traditional Newton-Raphson techniques. During the 2-D DIC the image is divided into circular subsets which are used to correlate spatial regions between the two images, the larger this subset the greater precision with which displacement or position can be found but at the cost of spatial resolution, features smaller than the subset size are unable to be resolved. A circular subset of radius 21 pixels was found to work well for the current implementation, being the smallest subset possible without producing noisy artefacts in the data, and has been used in the displacement, low rate and high rate tests. The comparison with the DaVIS software was performed at a 31 subset pixel size. 

\begin{figure}
        \centering
        \begin{subfigure}[b]{0.8\columnwidth}
                \includegraphics[width=\textwidth]{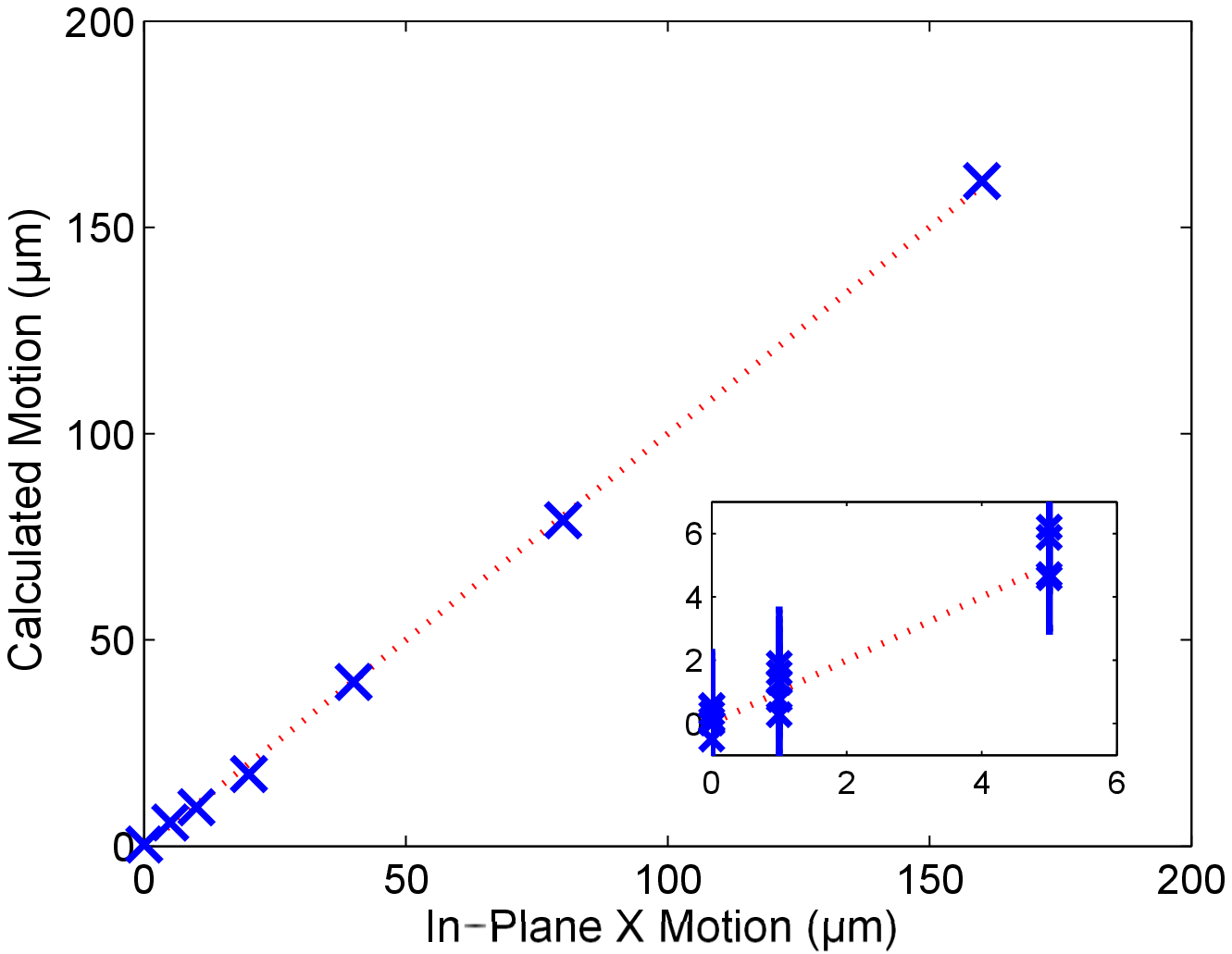}
                \caption{x-displacement}
        \end{subfigure}%
        ~ %add desired spacing between images, e. g. ~, \quad, \qquad, \hfill etc.
          
        \begin{subfigure}[b]{0.8\columnwidth}
                \includegraphics[width=\textwidth]{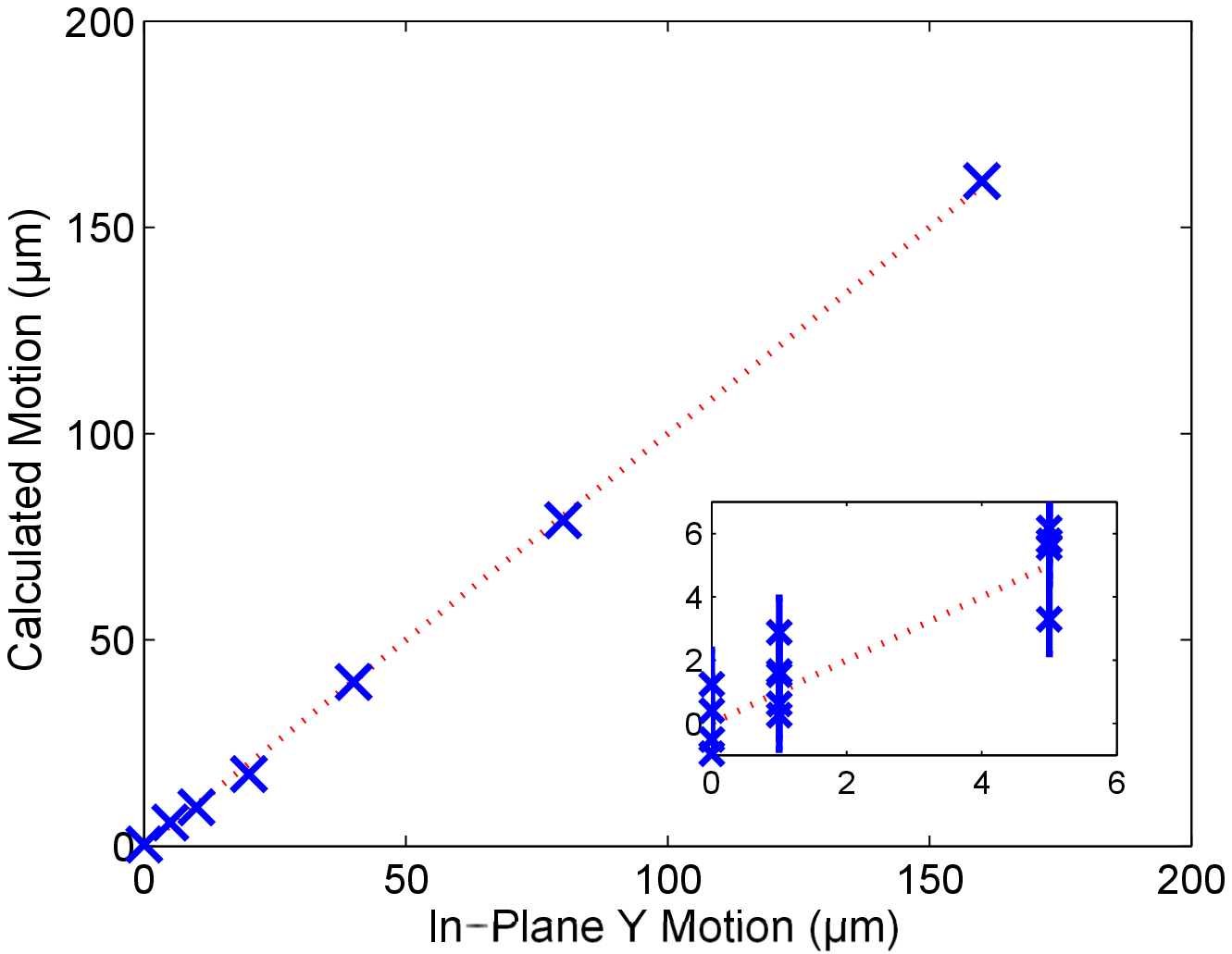}
                \caption{y-displacement}
        \end{subfigure}
        ~ %add desired spacing between images, e. g. ~, \quad, \qquad, \hfill etc.
          
        \begin{subfigure}[b]{0.8\columnwidth}
                \includegraphics[width=\textwidth]{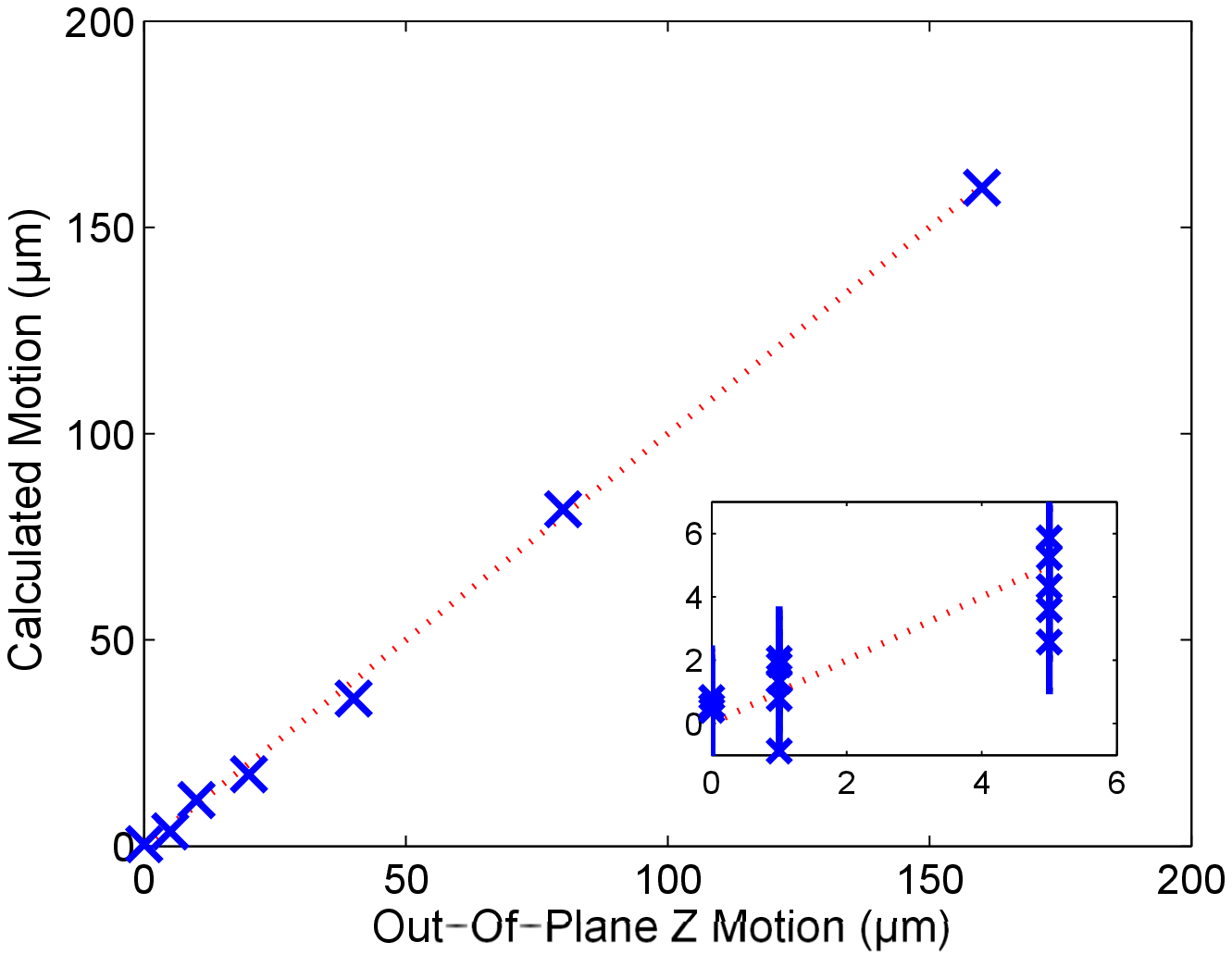}
                \caption{z-displacement}
        \end{subfigure}
        \caption{Displacement accuracy achieved with the phantom V7.3 with 240$\times$184 pixels. The results plotted represent an average over the spatial extent of the sample with error bars showing the standard deviation. The dotted line shows $y=x$.  The inset shows a close up of the 1 and 5 micron displacements where the error and fluctuations between measurements can be seen.}\label{fig:displp}
\end{figure}

Step 6 involves reconstructing the 3-D world from the 2-D information. By using the known camera parameters ($\mathbf{R_1,T_1,K_1,R_2,T_2,K_2}$) and equivalent pixel locations ($\mathbf{C_1,C_2}$) to solve the resultant simultaneous equations given by \eqref{camera1} and \eqref{camera2} the world pixel locations, $\mathbf{P_W}$, are obtained. In order to reconstruct a 3-D scene at a single time the 2-D DIC needs to be run between the left and right stereoscopic images only. However, obtaining 3-D displacements is more involved as it is not possible to perform 2-D DIC between the stereoscopic images at time $t_1$ and separately at time $t_0$ and subtract the results as image correspondence is lost, i.e. pixels at two differing times may not be looking at the same point in the 3-D space. Instead 2-D DIC must be performed between the images in as shown by the lines in the dashed box in Figure \ref{fig:workflow}. This enables the position of equivalent pixels in the each images at two differing times to be obtained and solves the sub-image correspondence problem, allowing the software to ascertain which parts of each image correspond to the same part in another \cite{PrenticeThesis}. This allows for subtraction of the world coordinates providing information on the 3-D displacement. A graphical user interface was developed in MATLAB which interfaced with the camera calibration toolbox and the Ncorr DIC system allowing the entire process to be carried out from within a single program. 

\section{Strain Calculation}
\label{sec:straincalculation}

Step 7 involves calculation of the strain map from the surface and displacement maps. Calculation of the full field strains is performed over a local volume element, centred on each reconstructed point and containing those points which lie within a predefined radius. Judicial choice of this radius is required to optimise the balance between accuracy and smoothness. The displacements are rotated into a local coordinate system with the $z$ axis locally normal to the surface. The displacement gradient tensor,

\begin{equation}
\mathbf{H}=\begin{bmatrix} \frac{\partial u_x}{\partial x} & \frac{\partial u_x}{\partial y} &\frac{\partial u_x}{\partial z} \\   \frac{\partial u_y}{\partial x} &\frac{\partial u_y}{\partial y} & \frac{\partial u_y}{\partial z}  \\    \frac{\partial u_z}{\partial x} & \frac{\partial u_z}{\partial y} & \frac{\partial u_z}{\partial z}\end{bmatrix}
\end{equation} can then be calculated through linear regression. The deformation gradient tensor ($\mathbf{F}=\mathbf{H}+\mathbf{I}$), Cauchy strain tensor ($\mathbf{C}=(\mathbf{F}^T\mathbf{F}$) and Lagrange strain tensor ($\mathbf{E}=\frac{1}{2}(\mathbf{C}-\mathbf{I})$) are defined in the usual way. It should be noted that the derivatives with respect to the surface normal direction, here defined as $z$, cannot be obtained from measurements of the surface displacement without further assumptions which leads to only the $E_{xx}$, $E_{yy}$, $E_{xy}$ and $E_{yx}$ components of the Lagrange strain tensor being defined, with the in-plane local directions {x} and {y} chosen in advance. %The major and minor strain values can be calculated from the eigenvalues of the resultant $2\times2$ matrix, $\epsilon=\sqrt{1+2\lambda}-1$. 

\section{Displacement Accuracy}
\label{sec:displ}

An initial demonstration of the accuracy of any DIC system is to replicate known linear motions both in- and out-of-plane. For this purpose we drove a flat target with the speckle pattern shown in the inset of Fig. \ref{fig:exp} distances of up to 160~$\mu$m in the x-, y- and z-directions, with the z-direction corresponding to out-of-plane displacement. Using Thorlabs Z825BV actuators the speckle pattern was driven to a specified distance, always moving in the direction of positive travel to avoid backlash. At each distance the sample was held stationary and an image taken. The same speckle pattern was used for each of the tests. Figure \ref{fig:displc} shows the displacements measured with the high resolution Canon set-up and the read-out from the linear actuators; the results show exceptional agreement between the calculated and actual displacements with an RMS error of 0.5~$\mu$m. In order to test the performance of the reconstruction algorithm during high rate experiments the Canon EOS 600D was replaced with a Phantom V7.3 high speed camera using just 240x184 pixels of the CCD. At this resolution the camera is capable of running at 50000 frames per second, the results are shown in Figure \ref{fig:displp} and demonstrate a RMS error of 1.6~$\mu$m. The RMS error was found to be similar, i.e. within 20\%, in each of the orthogonal directions.

However, it is not clear if the reduction in precision can be attributed to the reduced optical resolution, the applicability of the speckle pattern at this resolution or instability introduced through the cooling fan in the Phantom V7.3. To test for sources of vibration the displacement of subsequent images of a stationary target were measured and found to be 1.6~$\mu$m when taken 1~s apart and 1.5~$\mu$m when taken 0.2~ms apart, suggesting some component of the resolution is due to vibrational motion. This becomes more pronounced when running the Phantom with the full CCD area (800x600 pixels) at 6600 frames per second. In this case the displacement of subsequent images of a stationary target were found to be 1.5~$\mu$m when taken 1~s apart and 0.5~$\mu$m when taken 0.15~ms apart. This discrepancy suggests a source of vibration in the measurements with the Phantom 7.3. The lack of vibrational motion in the high speed data does not preclude the possibility that the source of error is from fluctuations in the target environment, the lack of such motion with the Canon is highly suggestive that the cause is internal to the Phantom. Indeed, the Phantom V7.3 uses a continuous cooling fan while newer models (e.g. v2511) contain a `Quiet Fans' mode which turns off the fans for a short period specifically for vibration-sensitive applications. 

It should be noted that for optimal performance the idealised speckle pattern should be modified to match the field of view and resolution \cite{suttonbook}, however this work utilised the same pattern in each case, therefore it may be possible to improve these results further. However, the achieved accuracy is consistently around 200 times greater than the optical resolution, comparable to other DIC systems \cite{singlecamera1}, and as such we would expect this to give marginal improvements.

\begin{figure*}
        \centering
        \begin{subfigure}[b]{0.5\columnwidth}
                \includegraphics[width=\textwidth]{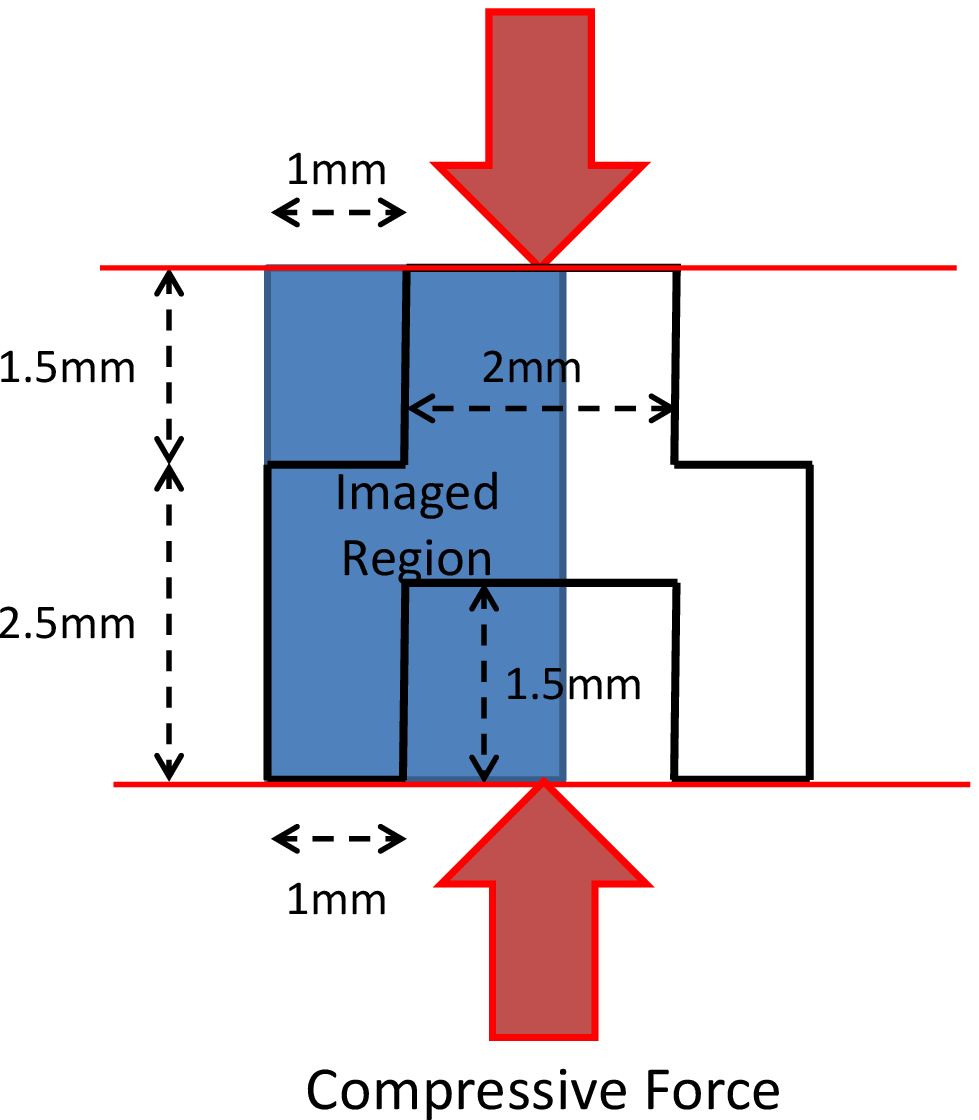}
                \caption{Top-hat sample}
                \label{fig:tophat} 
        \end{subfigure}%
        ~ %add desired spacing between images, e. g. ~, \quad, \qquad, \hfill etc.
        \begin{subfigure}[b]{0.5\columnwidth}
                \includegraphics[width=\textwidth]{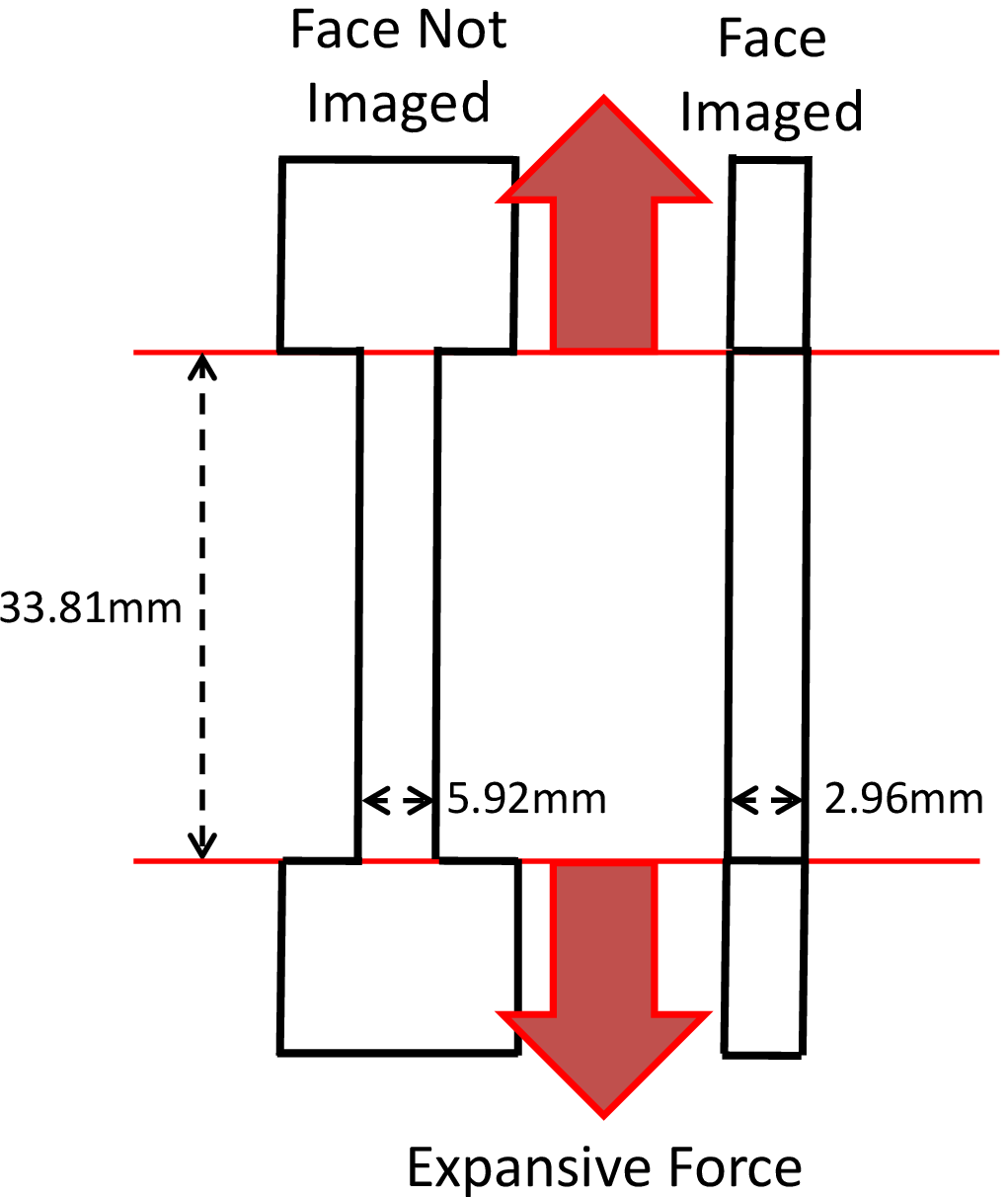}
                \caption{Dog-bone sample}
                \label{fig:dogbone} 
        \end{subfigure}
                \begin{subfigure}[b]{0.55\columnwidth}
                \includegraphics[width=\textwidth]{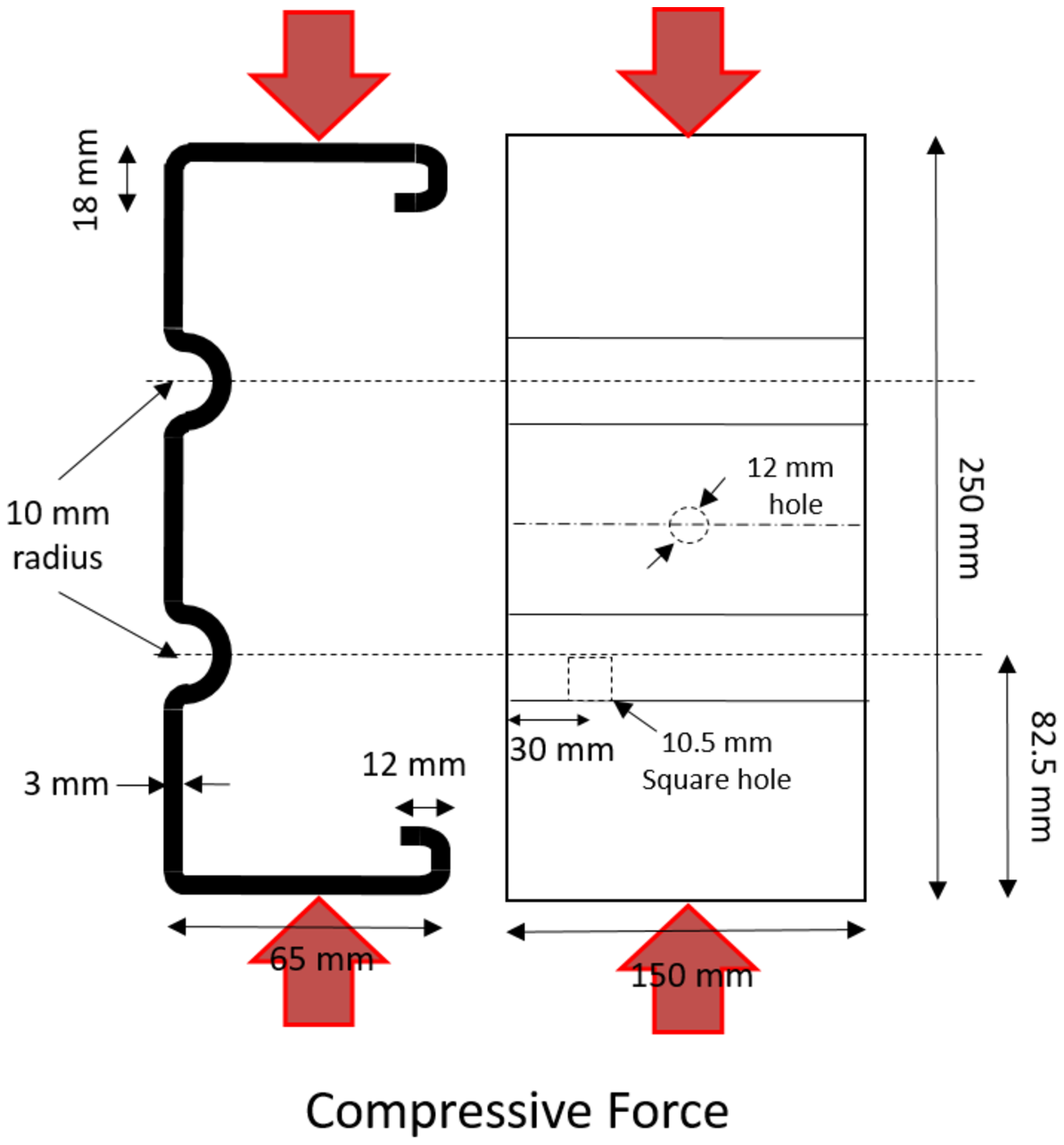}
                \caption{Steel C-section sample}
                \label{fig:daviscom} 
        \end{subfigure}
        ~ %add desired spacing between images, e. g. ~, \quad, \qquad, \hfill etc.
\caption{Dimensions of the samples used to test the system. (a) Top-hat sample of IMI834 Titanium alloy used in high resolution shear band test. Arrows indicate the direction of compressive force. (b) Dog-bone shaped sample of low-carbon steel used in a tension test on an Instron 5584 to demonstrate the out-of-plane z motion as the sample will begin to neck as it nears failure. (c) Dimensions of sample used in the comparison with the DaVIS commercial system. The specimen was a short length of cold-formed galvanised steel C-section cut from purlin used in a flooring system.}
\label{fig:samples}       % Give a unique label
\end{figure*}

\section{Comparison to DaVIS software}

\begin{figure}
        \centering
        
        \begin{subfigure}[b]{0.75\columnwidth}
                \includegraphics[width=\textwidth]{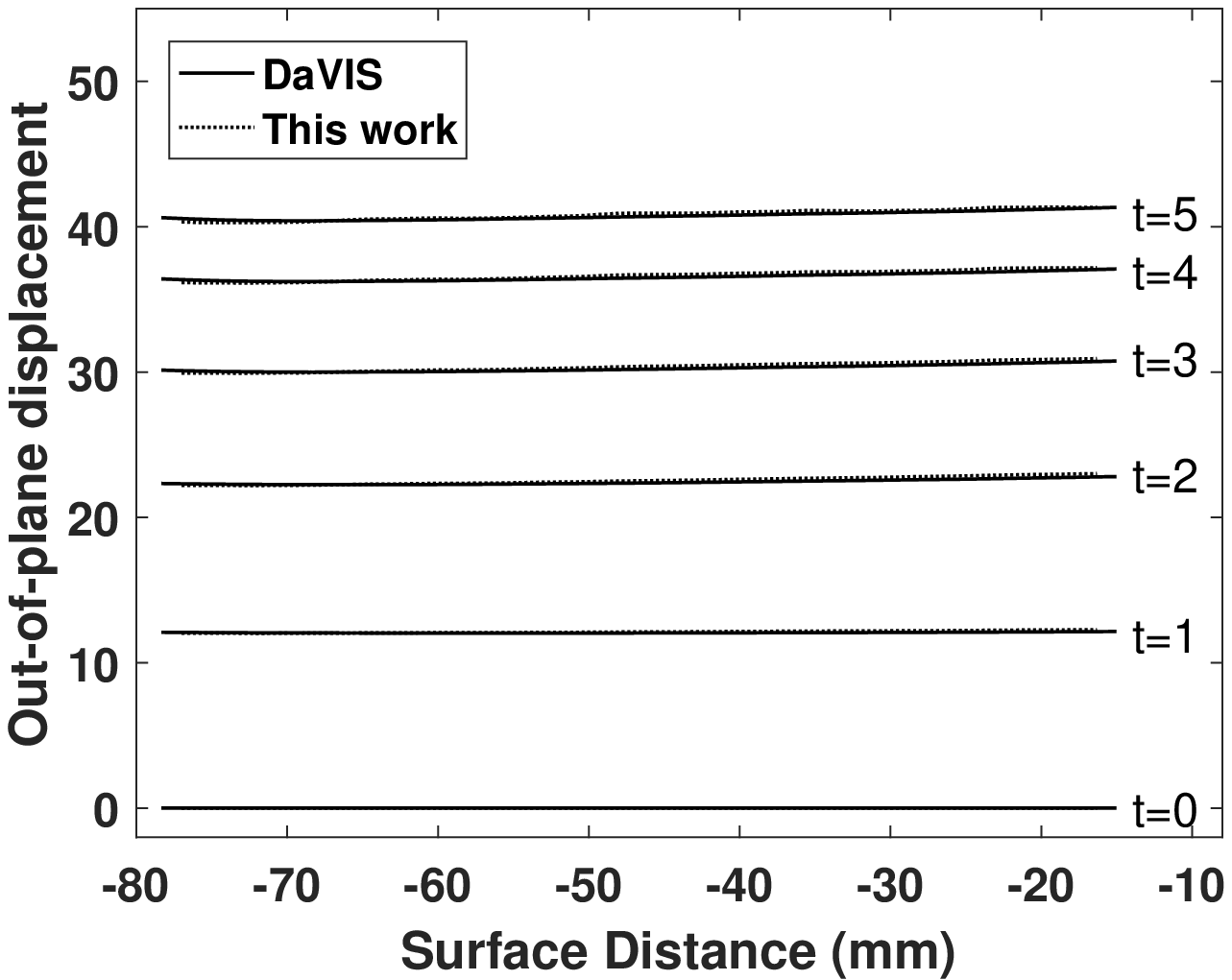}
                \caption{Out-of-plane displacement}
                                 \label{fig:davis_1}
        \end{subfigure}%
        
        \begin{subfigure}[b]{0.75\columnwidth}
                \includegraphics[width=\textwidth]{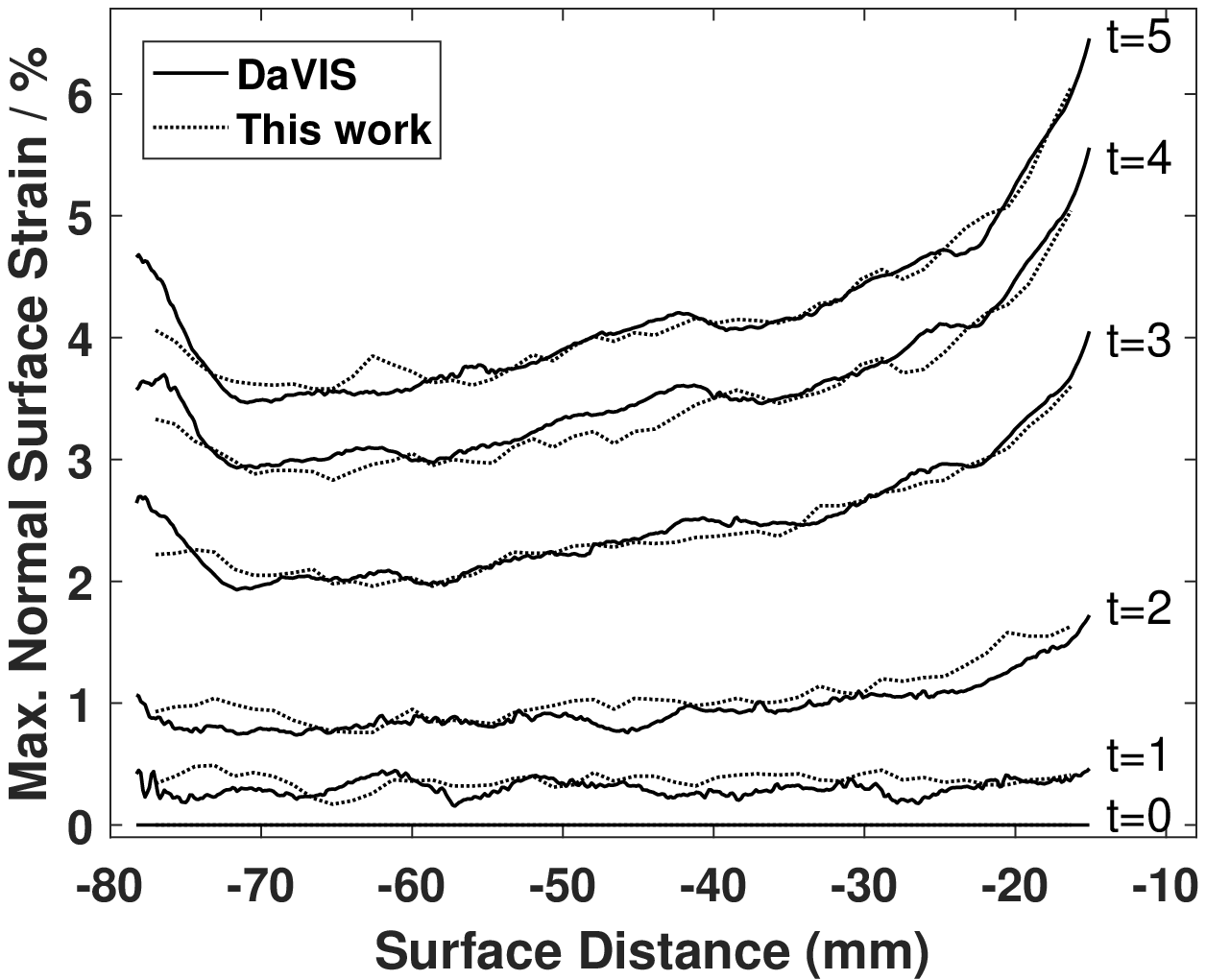}
                \caption{Max Normal Surface Strain}
                 \label{fig:davis_2}
        \end{subfigure}
        
          \begin{subfigure}[b]{0.7\columnwidth}
                \includegraphics[width=\textwidth]{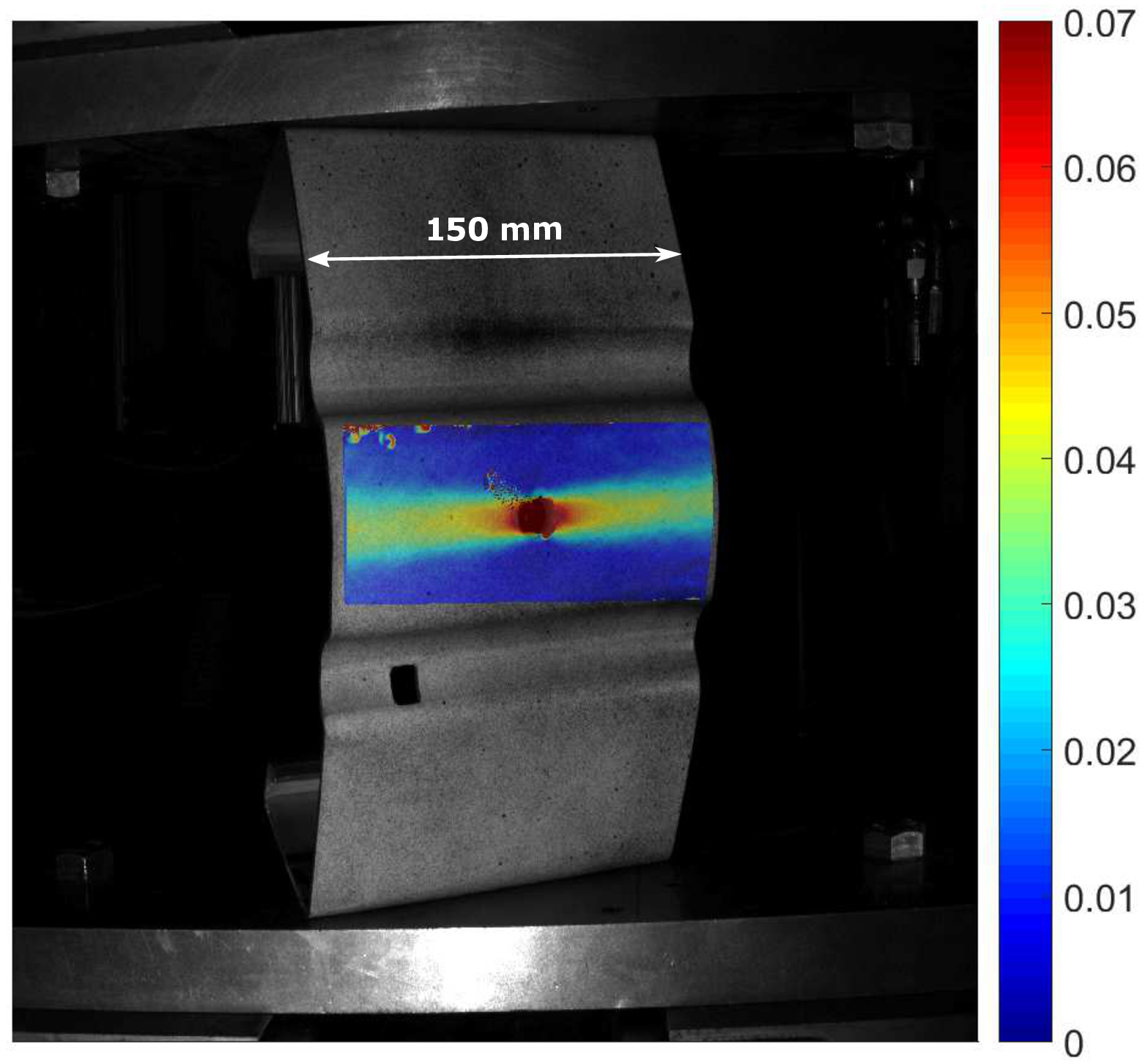}
                \caption{Max Normal Surface Strain}
                                 \label{fig:davis_3}
        \end{subfigure}%
        
\caption{Comparison of this 3D-DIC system with a commercial system. Shown are the out-of-plane displacements (a) and the maximum normal surface strain (b) calculated along the dashed-dotted line in Figure \ref{fig:daviscom} at 1 minute increments. Also shown is the maximum normal surface strain corresponding to t=5 minutes and projected onto the deformed image (c).}
\label{fig:davis}       % Give a unique label
\end{figure}

In order to demonstrate the applicability of the software we have performed a comparison of our system to a commercially available code, in this case the DaVIS software by LaVision\cite{davis}. A comparison test was performed on a short length of cold-formed galvanised steel C-section with the dimensions given in Figure \ref{fig:daviscom}. A two camera system compatible with the DaVIS software was used and were calibrated using their respective techniques, in the case of the DaVIS software this was a 3D calibration board, while our system was calibrated using the chequerboard technique. The same images were fed into both codes to remove ambiguity around illumination, speckle quality, camera resolution or lens quality. Speckle was applied using standard black spray paint over a white base coating to give maximum contrast, see Figure \ref{fig:davis_1}. Compression was applied to the sample at a constant velocity using an Instron SATEC with a 600 kN loadcell, at 10 mm/min. 

The results of the compression test are given in Figure \ref{fig:davis}. Figure \ref{fig:davis_1} shows the displacement of the sample along the dashed-dotted line in Figure \ref{fig:daviscom}, results are shown for times corresponding to 1 minute snapshots from the beginning of the compression. Due to the lack of a fiducial the results from the two codes have been shifted horizontally to find the closest agreement and found to match to within 200 $\mu$m. 

Comparison of the local strain calculations between the two systems was carried out by calculating the maximum normal surface strain across the same line out. The maximum normal surface strain was calculated from the eigenvalues of the Lagrange strain tensor. All points within a 1 mm radius sphere are used to calculate each strain value. Results are shown in Fig. \ref{fig:davis_2}. The two techniques agree to within 0.2\% strain. A projection of the maximum normal surface strain onto the image of the deformed sampled is shown in Figure \ref{fig:davis_3} for clarity. 

\section{Low Strain Rate Tests}
\label{sec:example}

In order to demonstrate the performance of the system a quasi-static compression test and a quasi-static tension test were performed. The first of these involved investigating the shear strain induced in millimeter sized top-hat specimens when subjected to a compressive load. This test utilised the Canon EOS with full resolution and is principally designed to demonstrate the high resolution aspect of the system. 

\begin{figure*}
        \centering
        \begin{subfigure}[c]{0.6\columnwidth}
                \includegraphics[width=\textwidth]{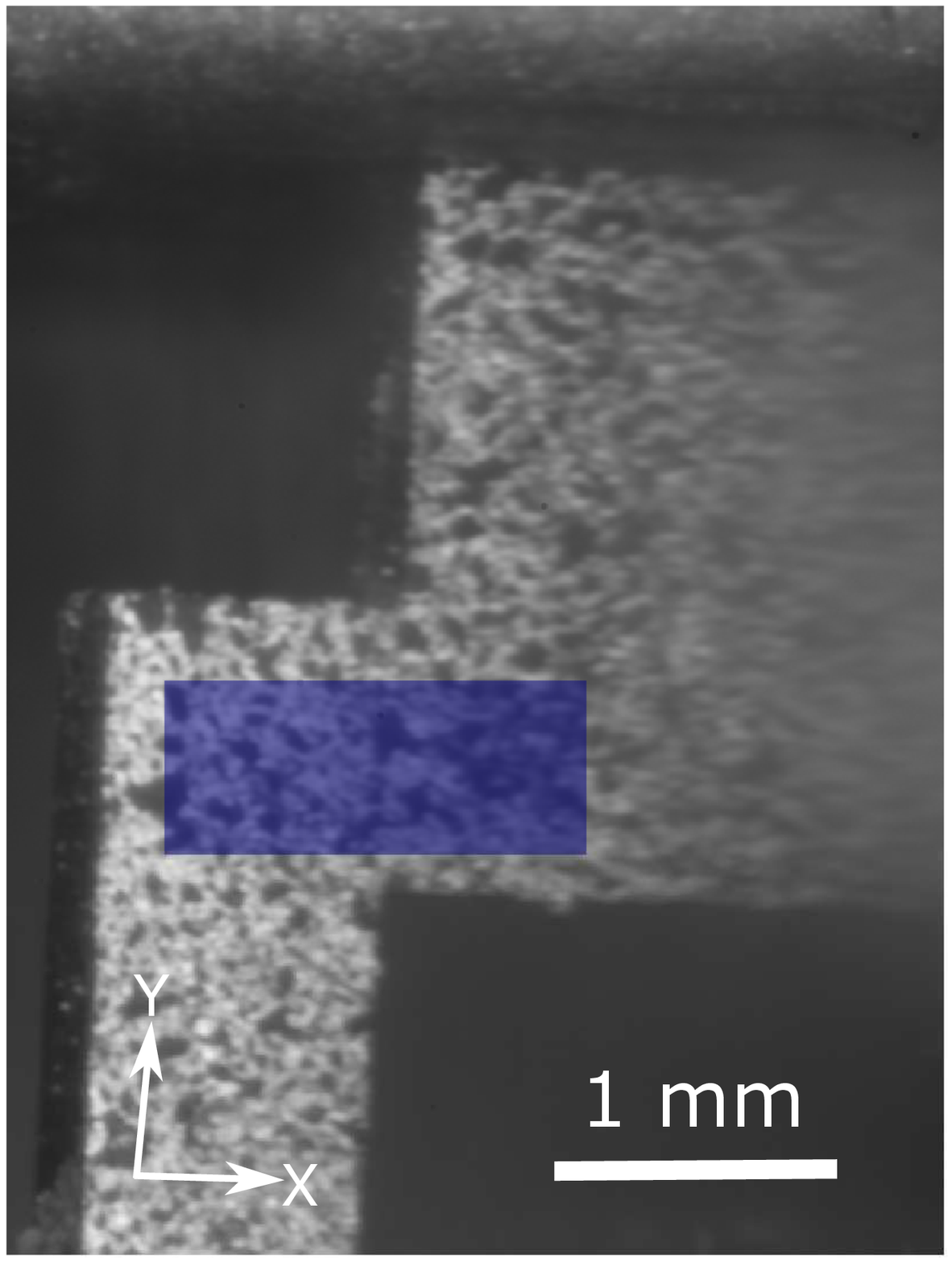}
                \caption{0\% global strain}
        \end{subfigure}\hspace{0.02\columnwidth}%
          \begin{subfigure}[c]{0.6\columnwidth}
                \includegraphics[width=\textwidth]{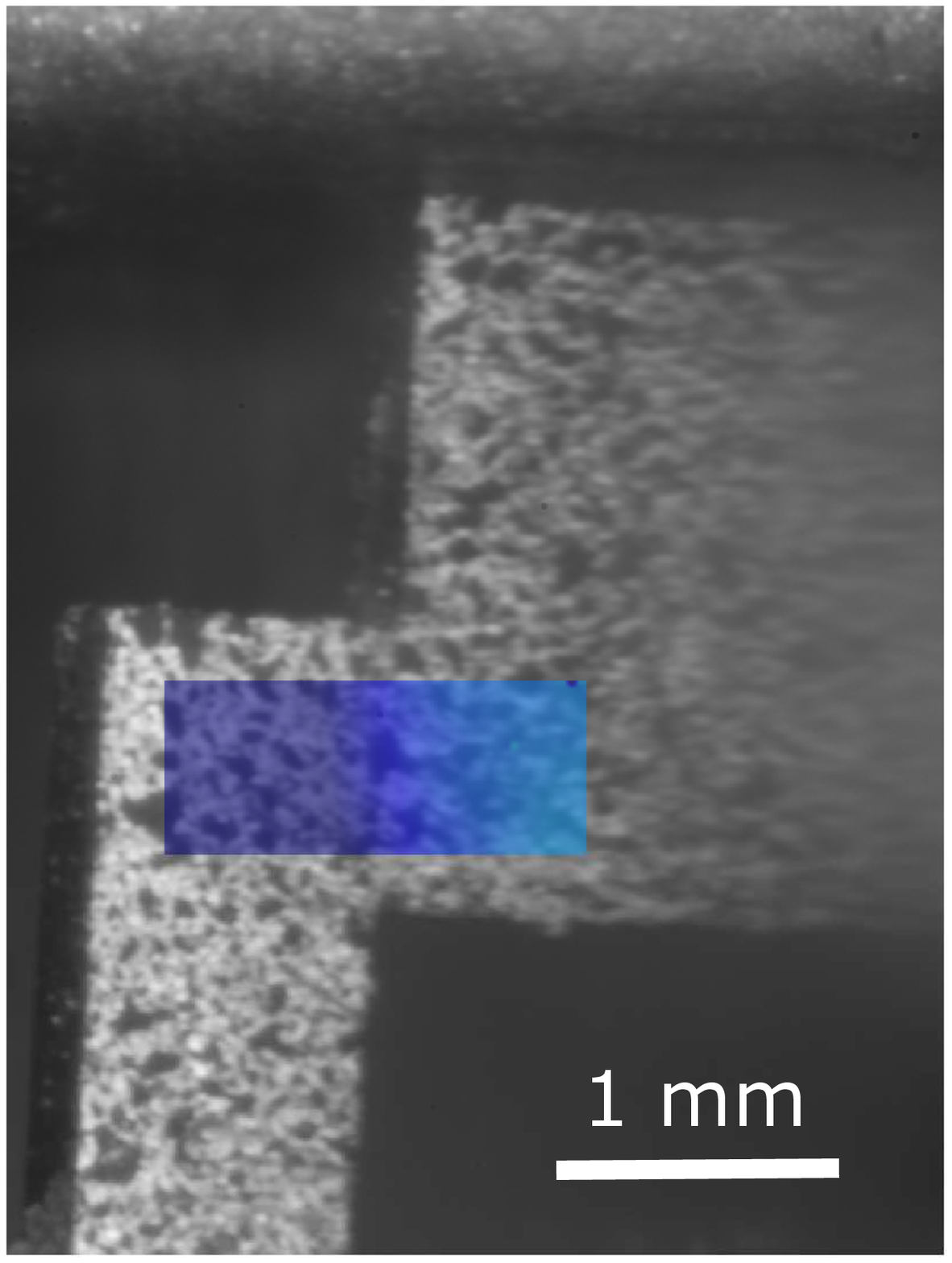}
                \caption{4.5\% global strain}
        \end{subfigure}\hspace{0.02\columnwidth}%
        \begin{subfigure}[c]{0.6\columnwidth}
                \includegraphics[width=\textwidth]{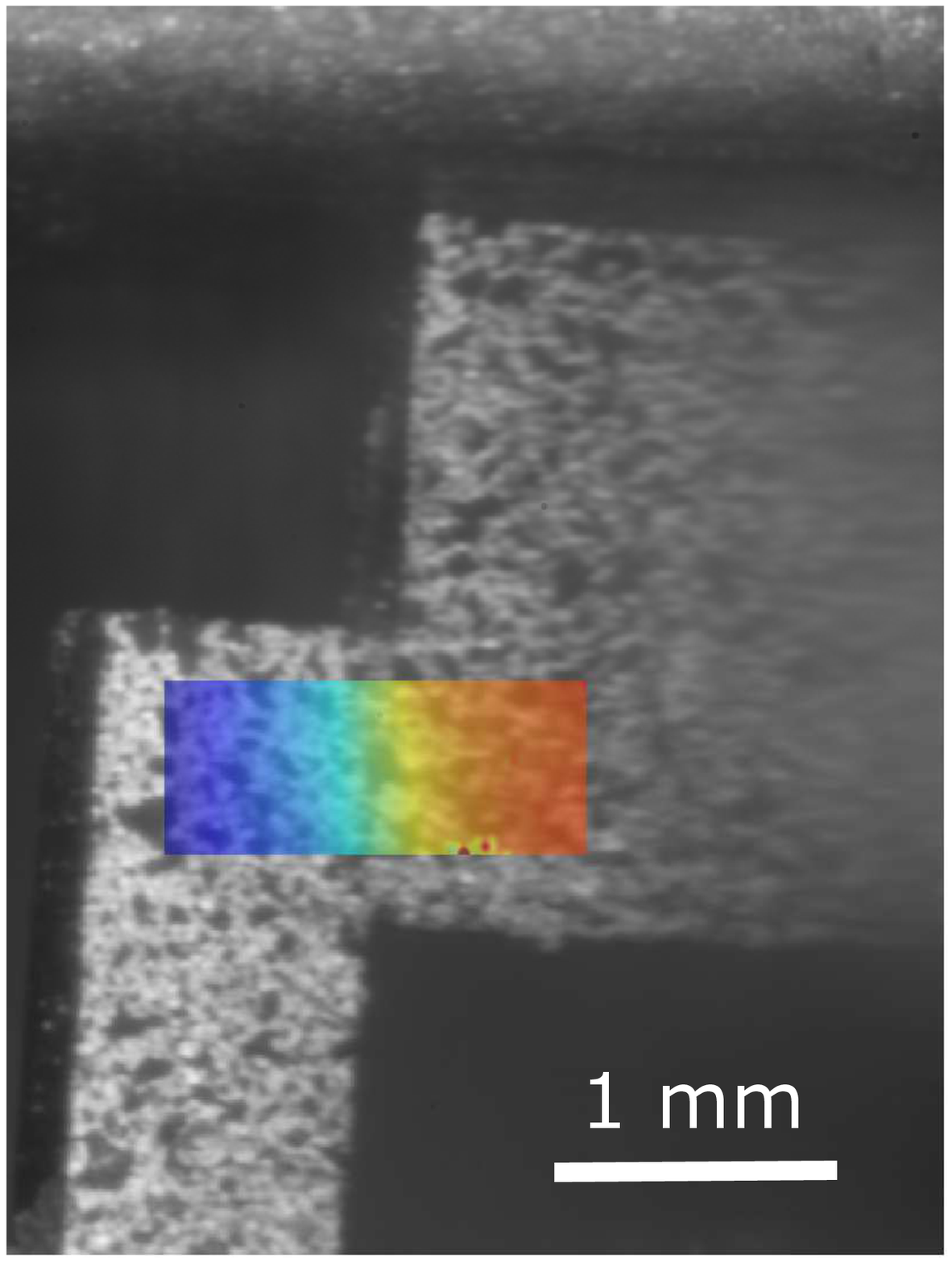}
                \caption{6.4\% global strain}
        \end{subfigure}
             
        \begin{subfigure}[c]{0.6\columnwidth}
                \includegraphics[width=\textwidth]{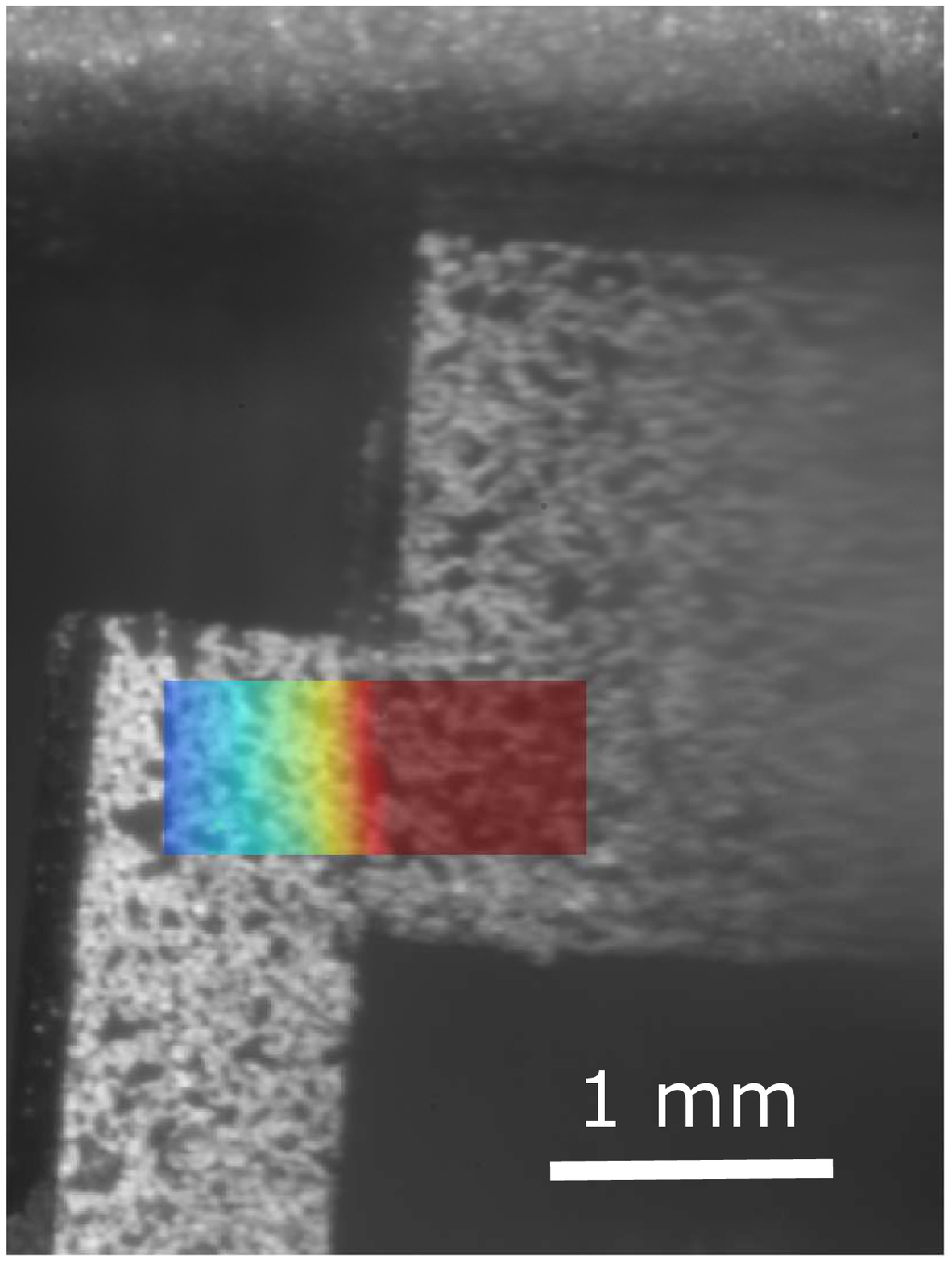}
                \caption{7.2\% global strain}
        \end{subfigure}\hspace{-0.02\columnwidth}%
        \begin{subfigure}[c]{0.2\columnwidth}
                \includegraphics[width=\textwidth]{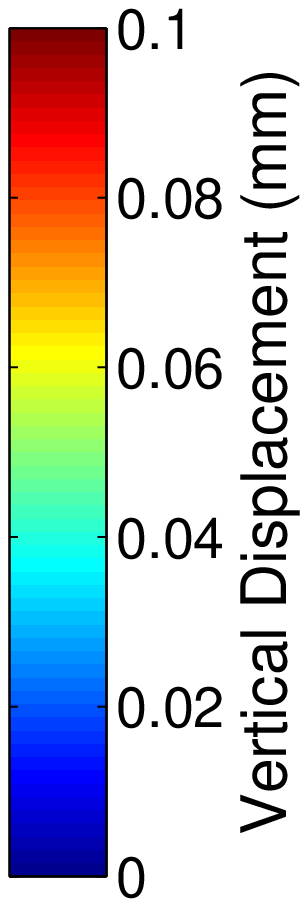}
        \end{subfigure}\hspace{0.02\columnwidth}%
        \begin{subfigure}[c]{1.04\columnwidth}
        \centering
                \includegraphics[width=0.88\textwidth]{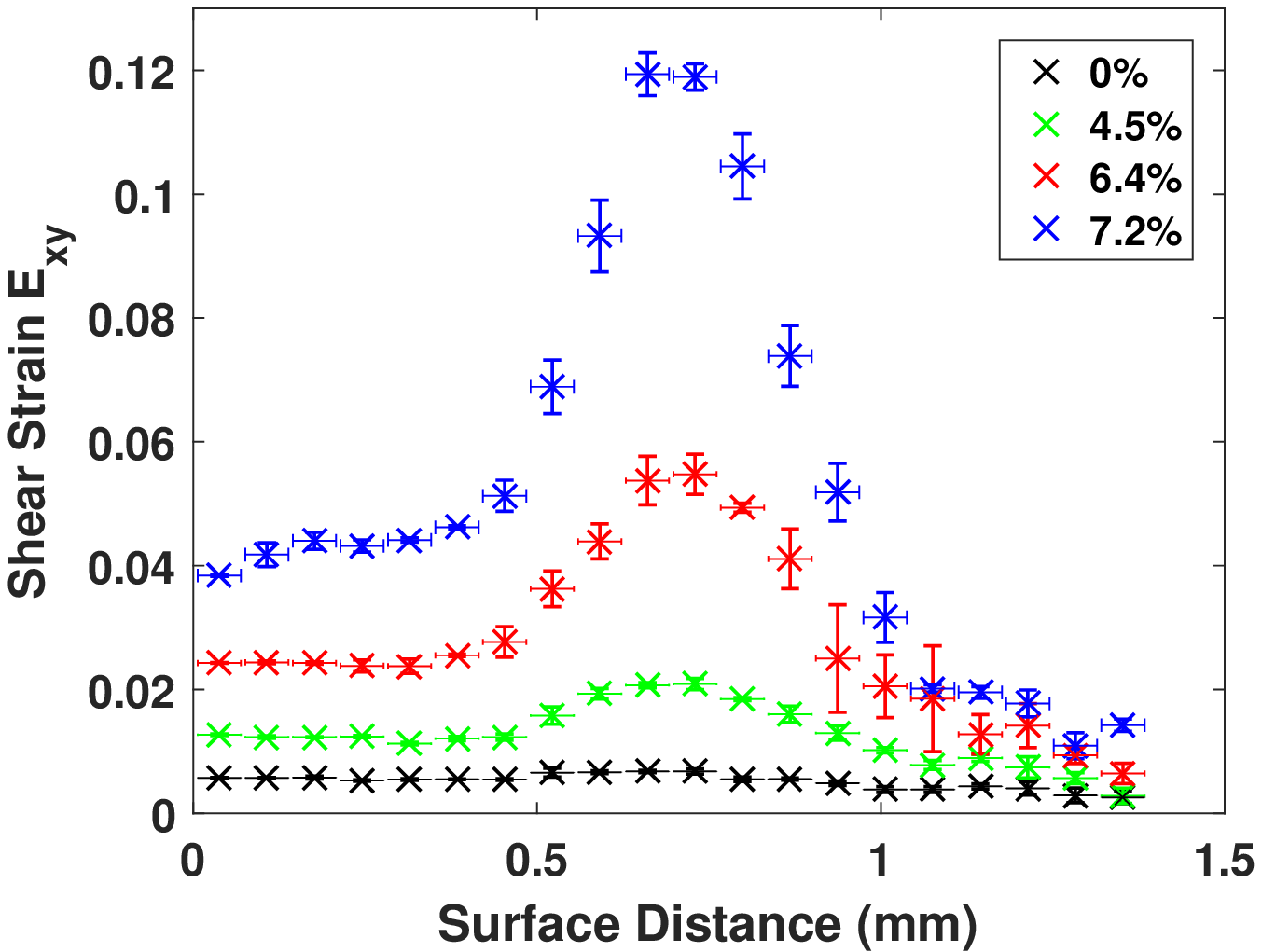}
                \caption{Shear strain observed in the sample.}
        \end{subfigure}%
        
\caption{Images of a top-hat sample under compression at global strains of  (a) 0\%, (b), 4.5\%, (c) 6.4\% and (d) 7.2\%, overlaid with a colour map showing vertical displacement. Figure (e) shows the average shear strain across the highlighted area of the sample.}
\label{fig:shear}       % Give a unique label
\end{figure*}

A flat faced top-hat shaped sample with dimensions given in Figure \ref{fig:tophat} was used. The samples were made from near-alpha titanium alloy IMI834, where high rate material behaviour has important relevance with regards to aero-engine discs, and in particular shear-band formation \cite{aerospace}. The characteristic shape and size of the sample leads to a small region of high shear strain localised in the region where the top-hat leg intersects the body \cite{tophat1,tophat2,tophat3,bdodd}. A compressive force was applied to the top of the sample while the legs of the sample were held still against an anvil. Figure \ref{fig:shear}a-d shows an overlay of the calculated y-displacement (that is displacement in the vertical direction) plotted against global strain, where global strain is defined as $\Delta{L}/L$ with $\Delta{L}$ sample compression and $L$ the initial length. Figure \ref{fig:shear}e shows a spatial average of the shear strain across the high shear region. Here, the shear strain is defined as $E_{xy}$, where $\mathbf{E}$ is the Lagrange strain tensor defined above and x and y correspond to local, in-plane directions shown in Figure \ref{fig:shear}a. As the global compressive strain is increased a shear band can be seen to develop, the width of this shear band, approximately $400~{\mu}$m, necessitates a high resolution DIC system which in this case is achieved through the use of the K2 lens. Towards the end of the test the shear strain increases further to a maximum observed shear strain of 0.12 and ultimately leads to sample failure. 

\begin{figure*}
        \centering
        \begin{subfigure}[c]{0.45\columnwidth}
                \includegraphics[width=\textwidth]{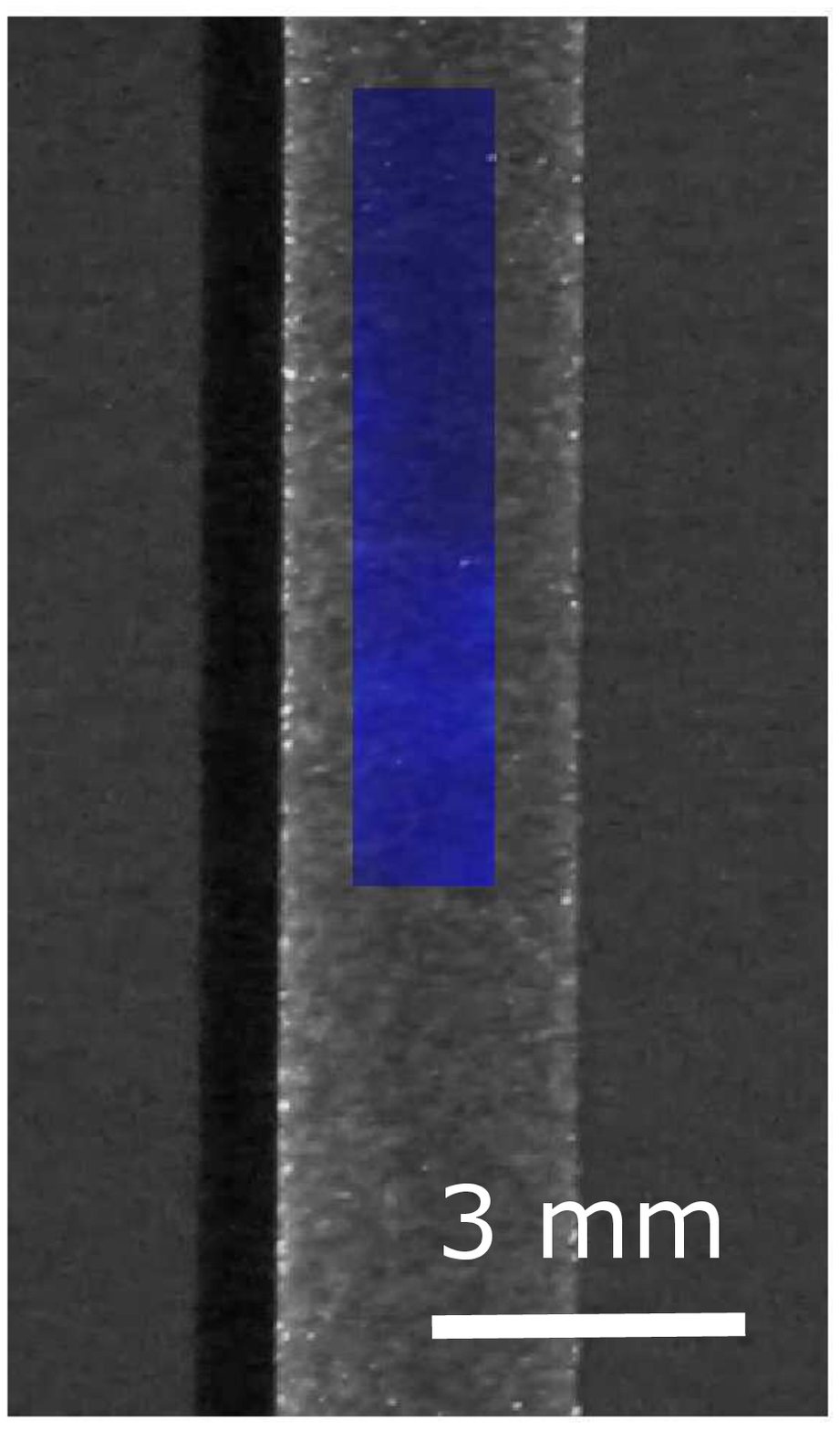}
                \caption{\\11.24 mm  Elongation}
        \end{subfigure}\hspace{-0.03\columnwidth}%
        \begin{subfigure}[c]{0.45\columnwidth}
                \includegraphics[width=\textwidth]{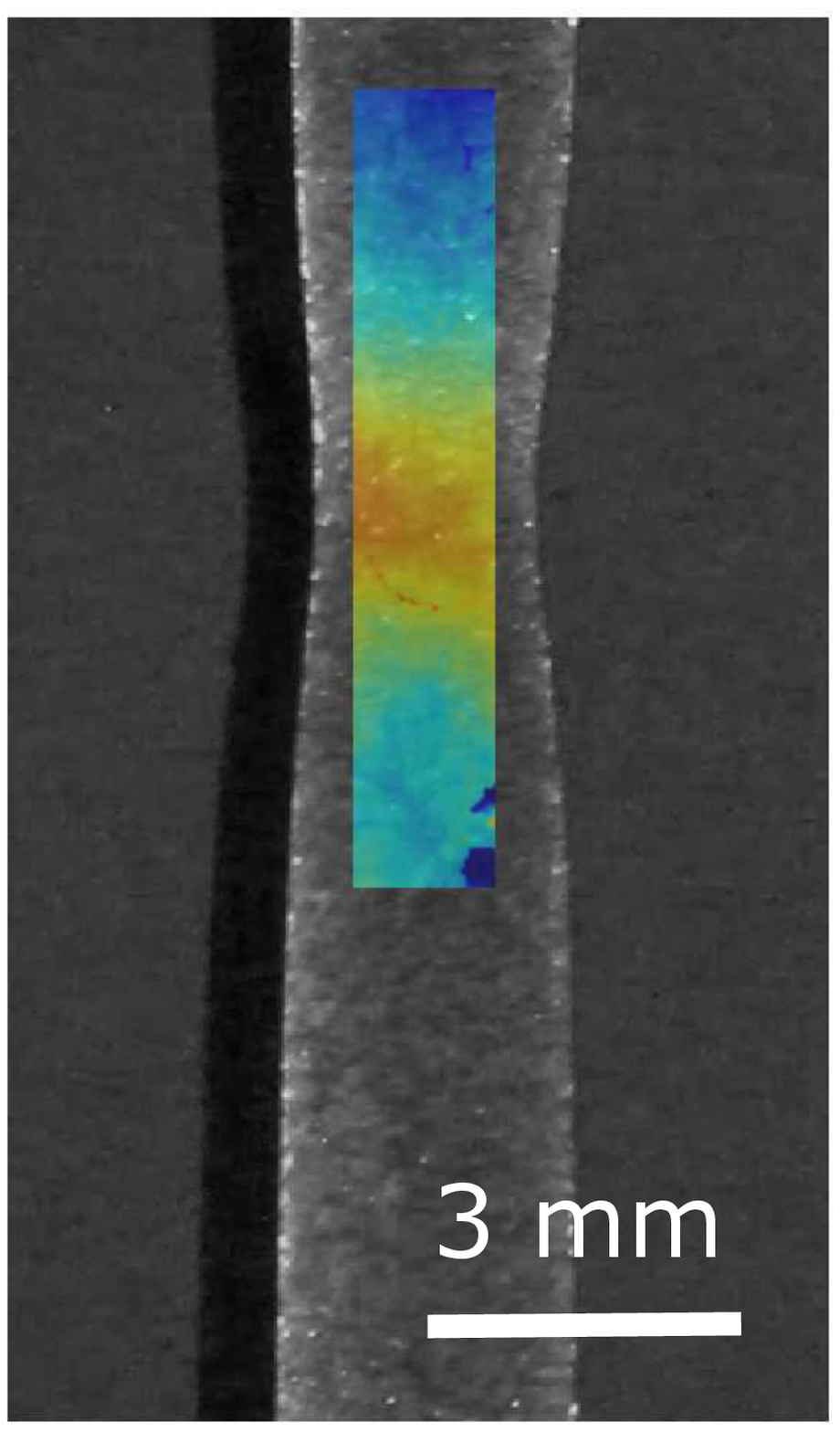}
                \caption{\\13.24 mm  Elongation}
        \end{subfigure}\hspace{-0.03\columnwidth}%
        \begin{subfigure}[c]{0.45\columnwidth}
                \includegraphics[width=\textwidth]{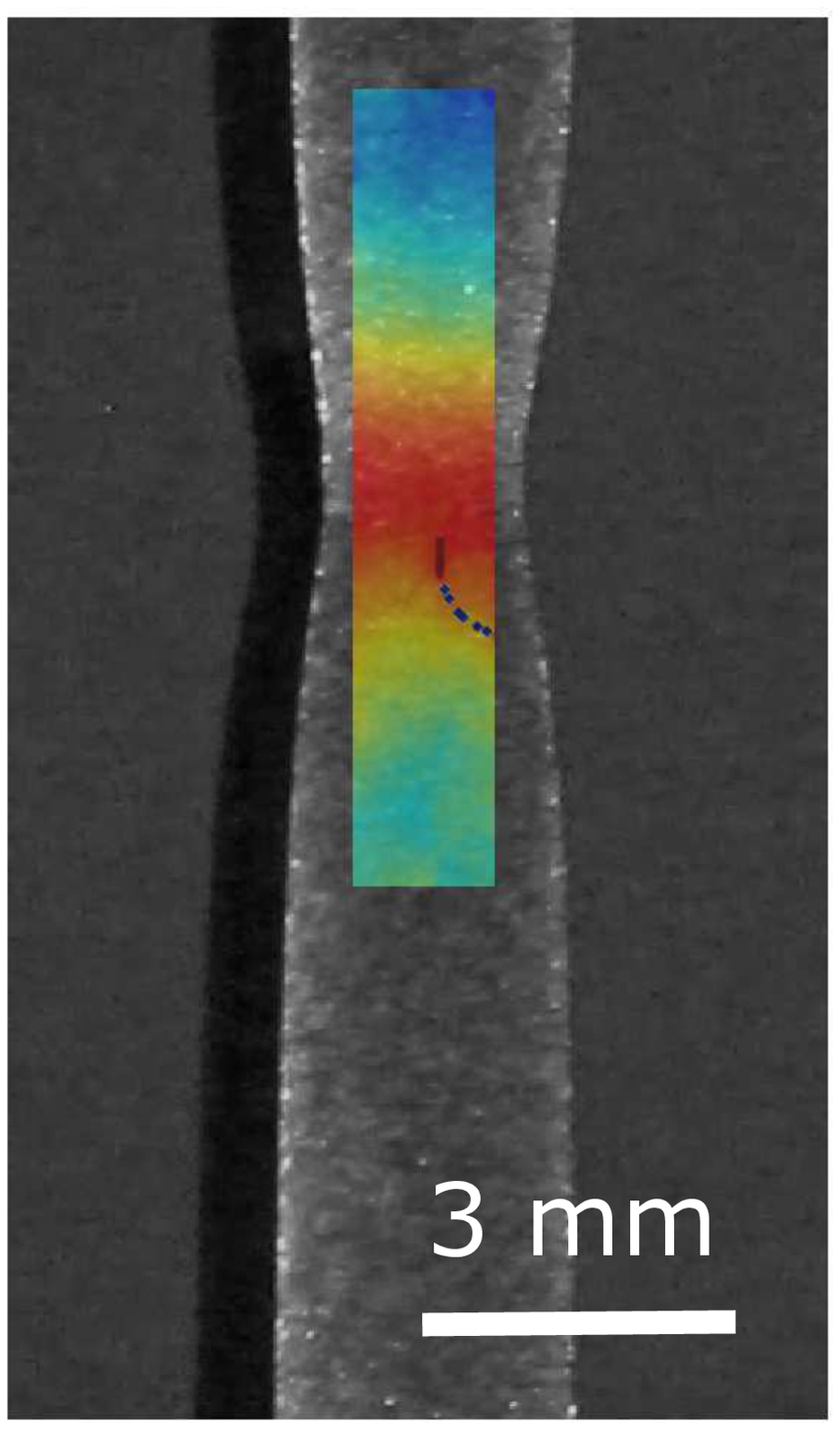}
                \caption{\\13.44 mm Elongation}
        \end{subfigure}\hspace{-0.03\columnwidth}%
        \begin{subfigure}[c]{0.45\columnwidth}
                \includegraphics[width=\textwidth]{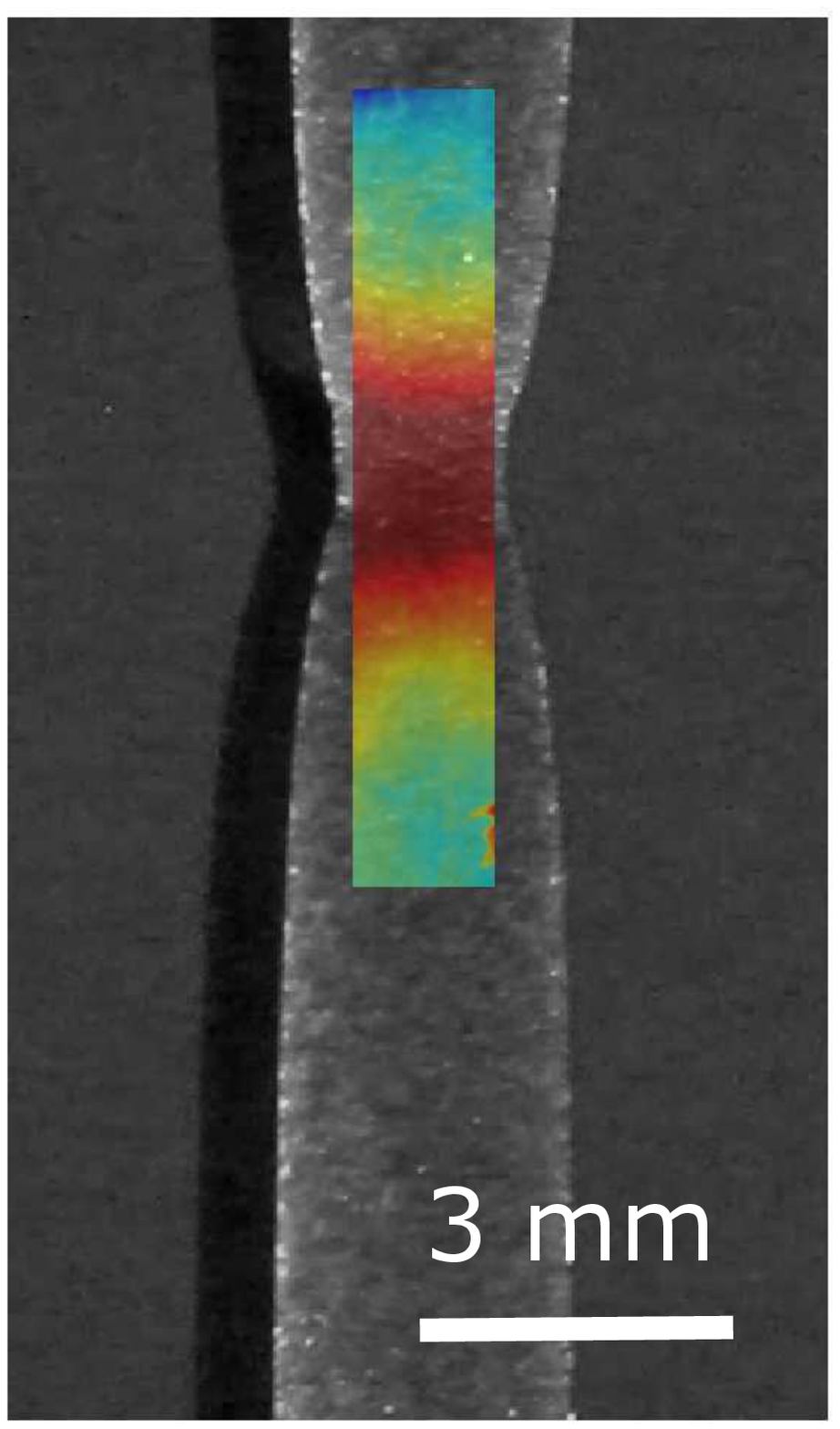}
                \caption{\\13.64 mm Elongation}
        \end{subfigure}\hspace{-0.02\columnwidth}%
        \begin{subfigure}[c]{0.2\columnwidth}
                \includegraphics[width=\textwidth]{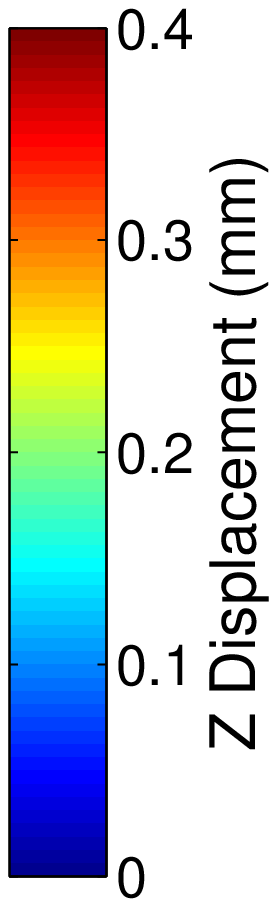}
        \end{subfigure}\hspace{0.02\columnwidth}%
        \begin{subfigure}[c]{0.45\columnwidth}
                \includegraphics[width=\textwidth]{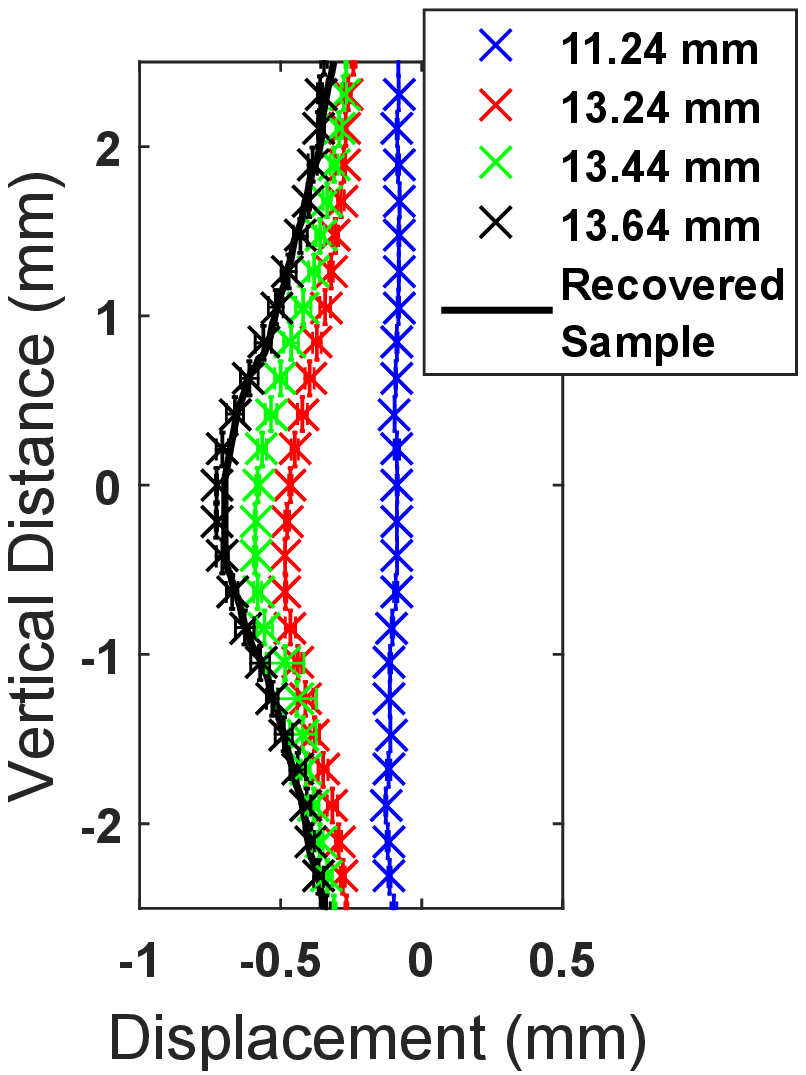}
                \caption{Sample Displacement}
        \end{subfigure}\hspace{0.05\columnwidth}%
                \begin{subfigure}[c]{0.7\columnwidth}
                \includegraphics[width=\textwidth]{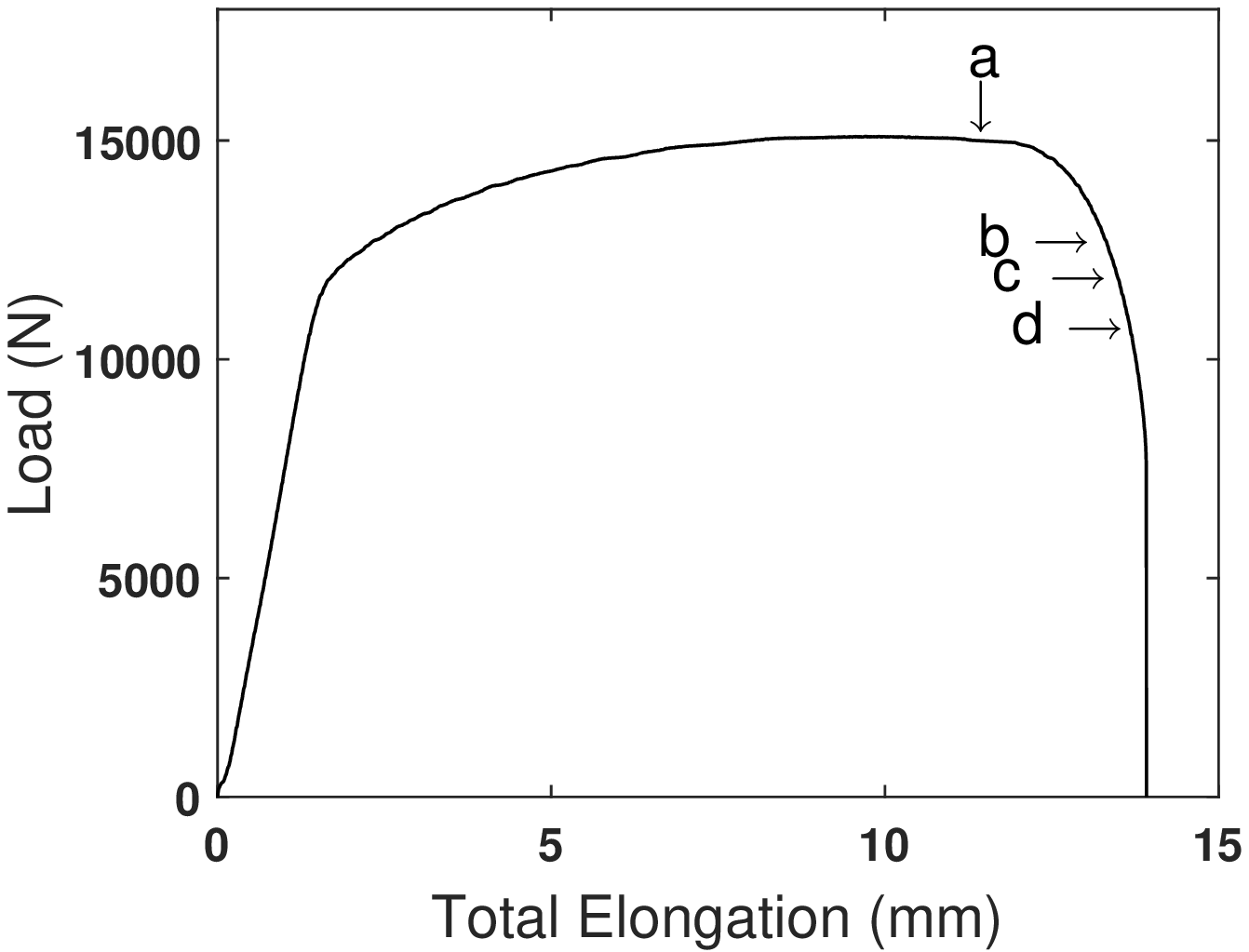}
                \caption{Stress-strain curve}
        \end{subfigure}
\caption{Evolution of tensile specimen subjected to a 100 kN load at a constant 1 mm/min. Shown are samples with a total elongation of (a) 11.24 mm, (b) 13.24mm, (c) 13.44 mm and (d) 13.64 mm. Figure (e) shows the out-of-plane surface position averaged across the highlighted region. Data is shown for four times chosen to cover the largest out-of-plane motion. Error bars representing the standard deviation across the width of the sample are smaller than the width of the markers. Figure (f) shows the corresponding stress curve. The location of the images (a)-(d) has been marked.  }
\label{fig:tension}       % Give a unique label
\end{figure*}

In order to test the 3-D aspect of the DIC system a quasti-static tension test was used. A dog-bone shaped sample of low-carbon steel was chosen due to its tendency to neck under tension, a schematic of the sample is given Figure \ref{fig:dogbone}. The sample was 33.81~mm long with a depth of 5.92~mm and width of 2.96~mm. The DIC system was set-up in order to image the thinner 2.96~mm face, this was chosen to maximise the out-of-plane motion during the test. In order to have a field-of-view that encompasses the full length of the sample the additional 4X magnification used throughout this work was removed. In this case it was found the surface roughness of the sample was such that a speckle pattern did not need to be applied. Tension was applied to the sample at a constant velocity using an Instron 5584, with a 100 kN loadcell, at 1 mm/min. Images of the resulting necking are shown in Figure \ref{fig:tension}, the out-of-plane motion averaged over the highlighted area is given in the insert. Each measurement contains less than $50~\mu$m displacement variation across the sample width. 

\section{High Strain Rate Tests}
\label{sec:highrate}

\begin{figure}
        \centering
   \includegraphics[width=\columnwidth]{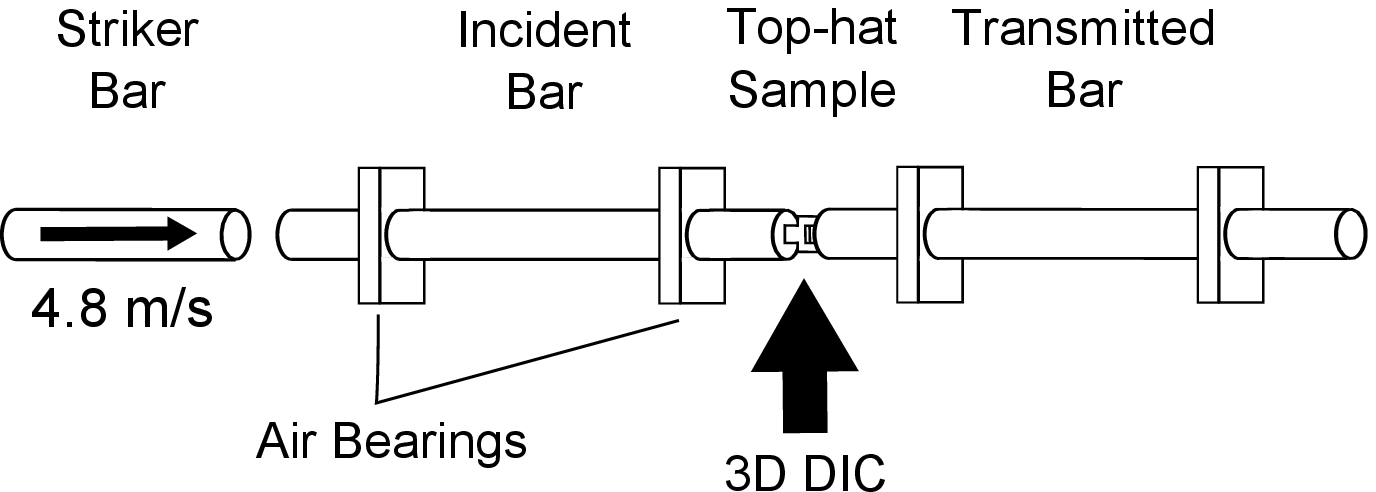}
        \caption{Schematic diagram of the Split-Hopkinson Pressure Bar used to demonstrate the high-strain aspect of the set-up.}\label{fig:shpb}
\end{figure}

The experimental set-up discussed above has been designed to offer a field of view of a few millimeters with a spatial resolution on the order of a few microns. The system was specifically designed to study the spatial and temporal characteristics of adiabatic shear bands, a phenomena typically associated with high strain-rate deformation and attributed to poor heat dissipation, thermal softening and ultimately material failure. Understanding this mechanism for failure is of the utmost importance in many industrial applications \cite{aerospace}.   

To further demonstrate the ease-of-use of the system we used a Phantom 2511 high-rate camera capable of recording 120 000 frames per second (8.3 $\mu$s inter-frame time) at a resolution of $240\times896$ pixels to record a top-hat sample under high strain-rate compression. The high-rate loading was performed on a split-Hopkinson pressure bar \cite{shpb1,shpb2,shpb3} utilising 6 mm diameter maraging steel bars held in position with frictionless air bearings. A 40 cm long striker capable of supplying a 140~$\mu$s loading pulse is used to compress millimeter-sized metallic samples at strain rates ranging from $10^2$ to $10^4$ s$^{-1}$. A schematic of the set-up is shown in Figure \ref{fig:shpb}. 

\begin{figure}
        \centering
        \begin{subfigure}[c]{0.65\columnwidth}
                \includegraphics[width=\textwidth]{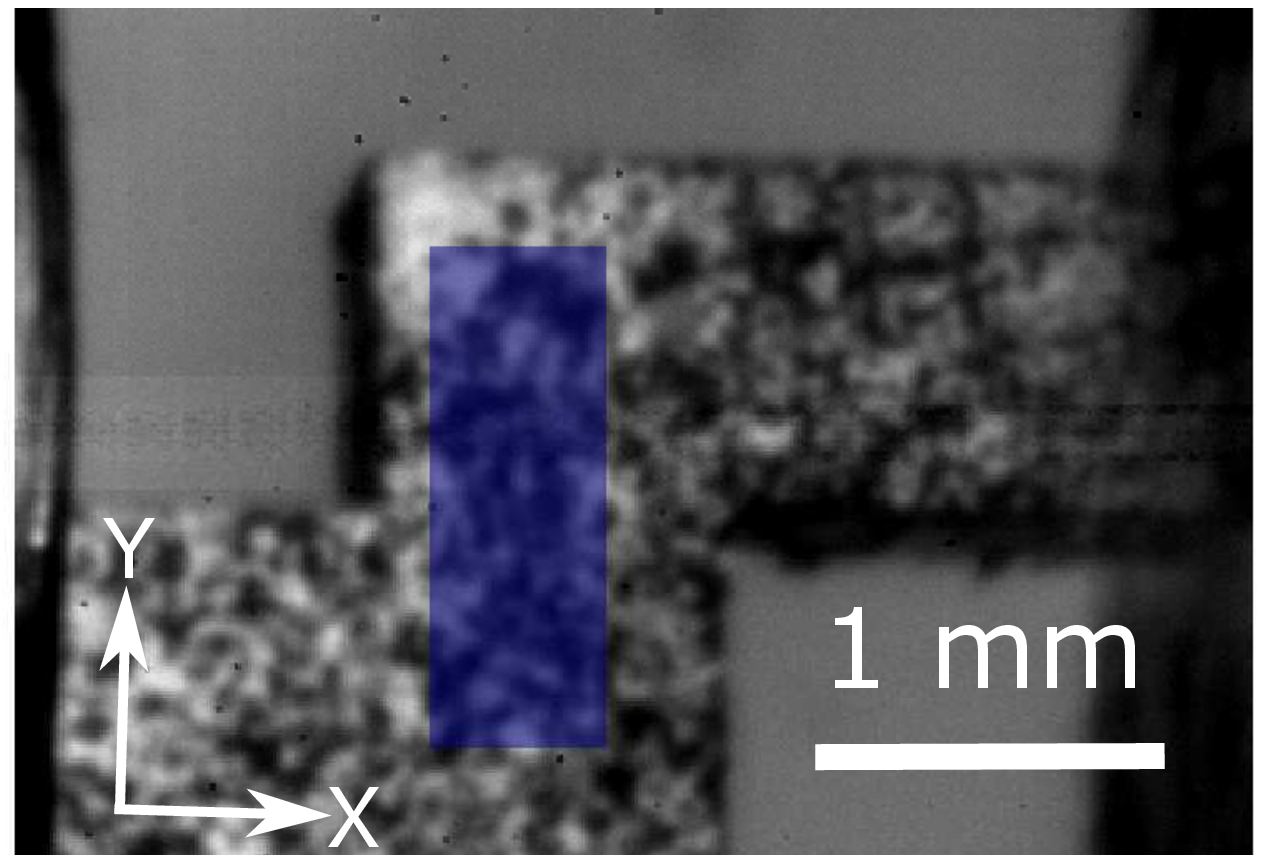}
                \caption{t = 0~$\mu$s}
        \end{subfigure}%
        
        \begin{subfigure}[c]{0.65\columnwidth}
                \includegraphics[width=\textwidth]{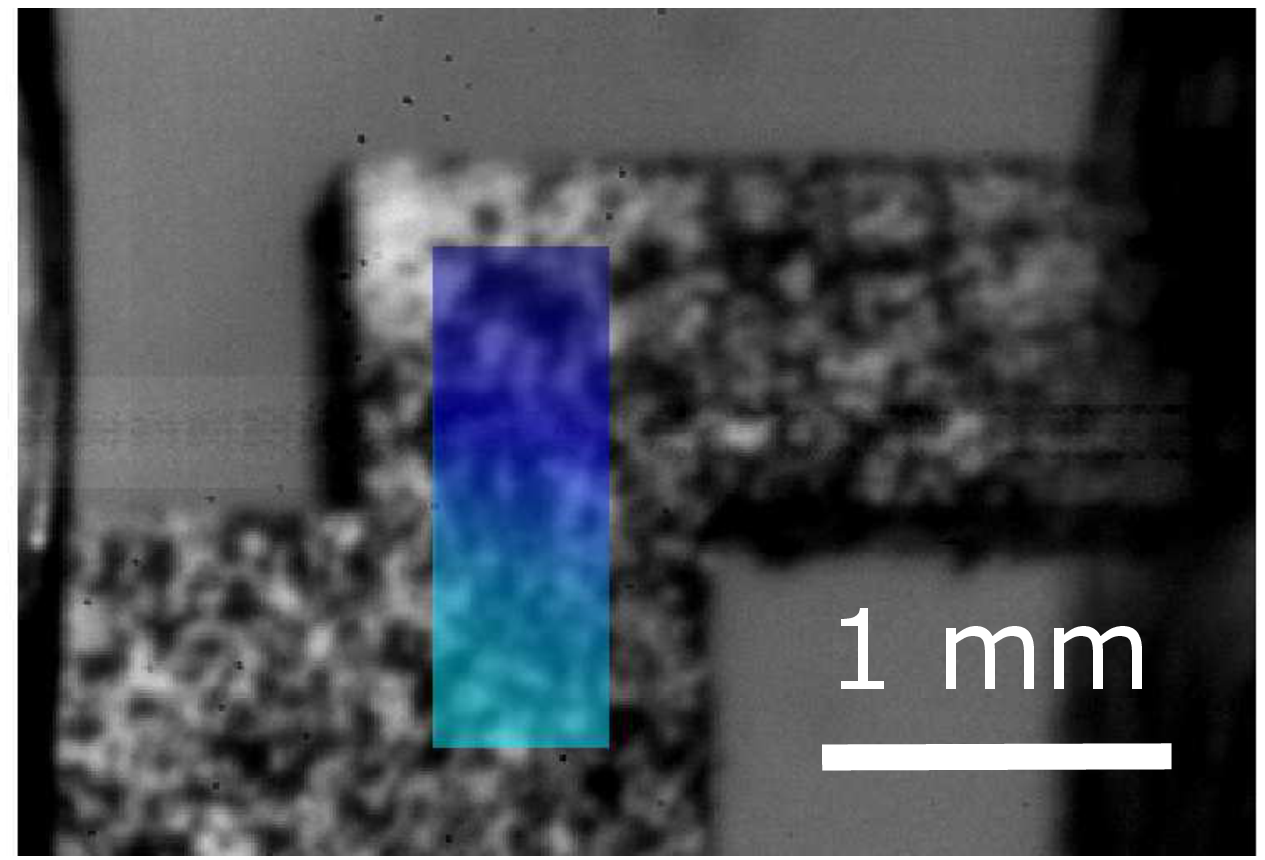}
                \caption{t = 16.6~$\mu$s}
        \end{subfigure}%
        
        \begin{subfigure}[c]{0.65\columnwidth}
                \includegraphics[width=\textwidth]{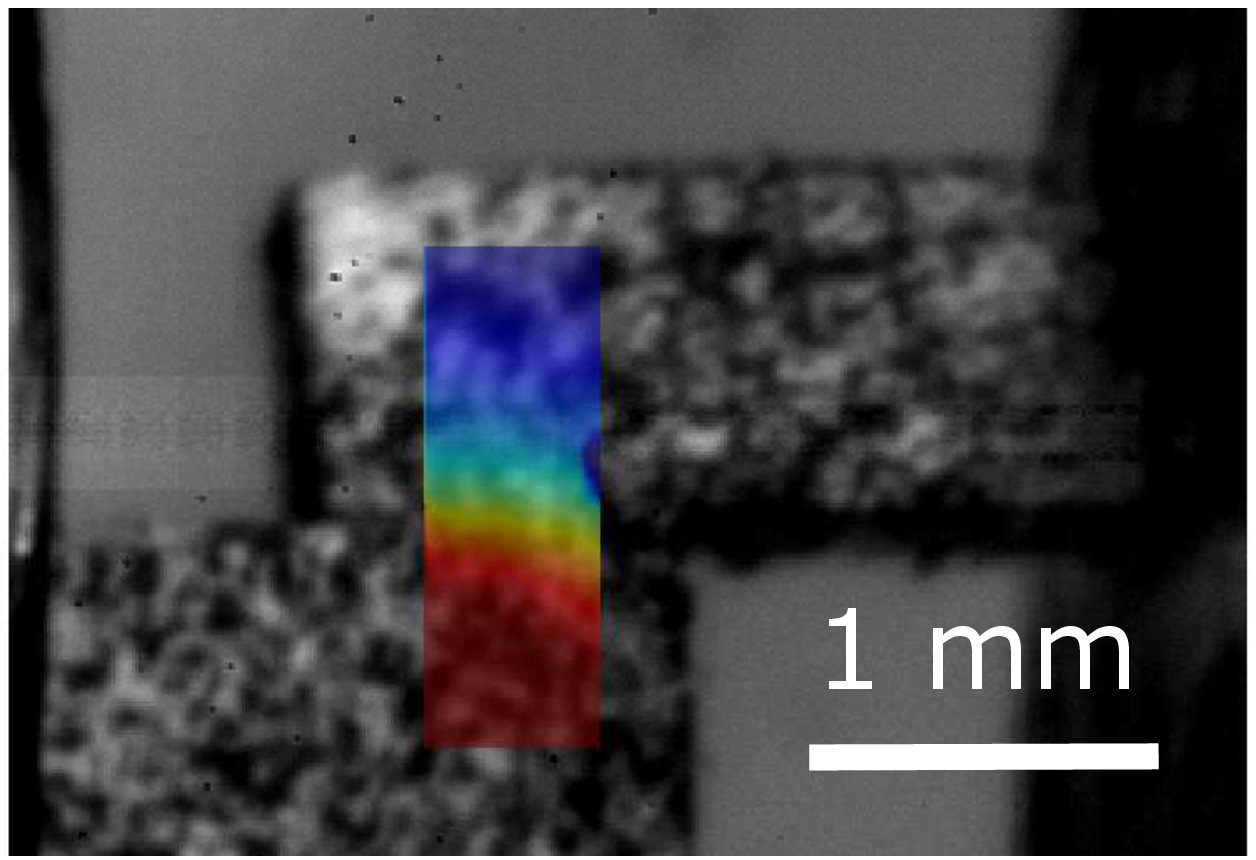}
                \caption{t = 33.3~$\mu$s}
                 \label{fig:highrate_last}
        \end{subfigure}
        
        \begin{subfigure}[c]{0.6\columnwidth}
                \includegraphics[width=\textwidth]{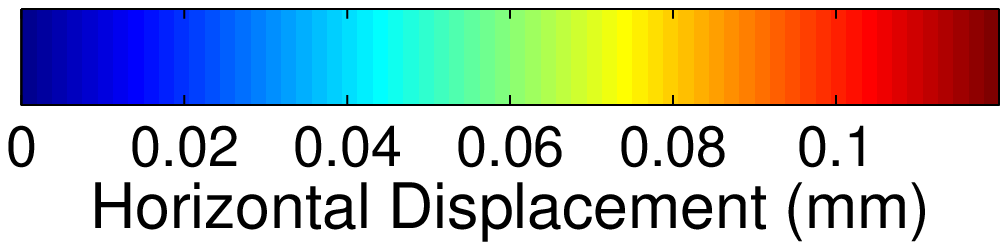}
        \end{subfigure}%
        
        \begin{subfigure}[c]{0.7\columnwidth}
        \centering
                \includegraphics[width=\textwidth]{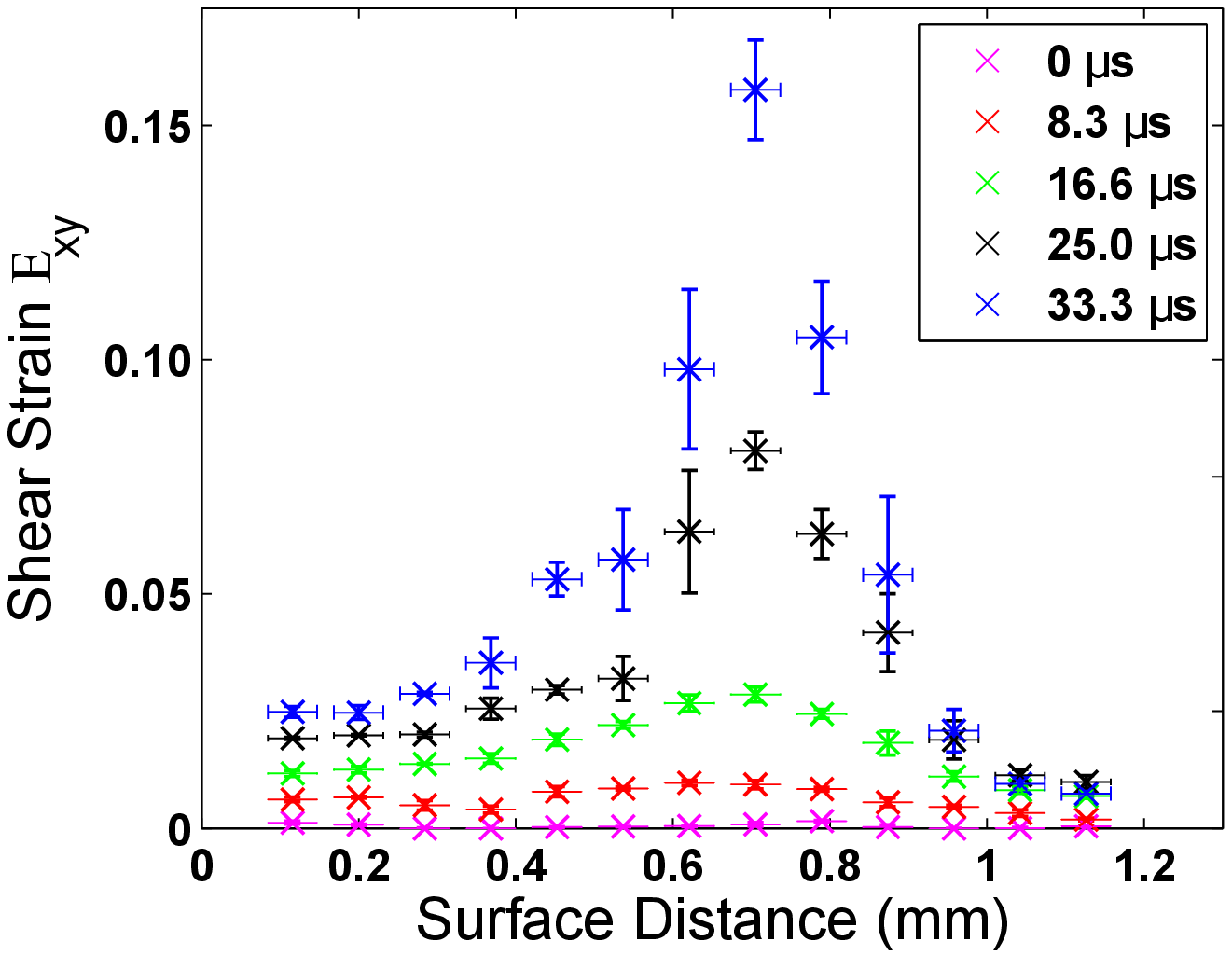}
                \caption{Shear strain observed in the sample.}
        \end{subfigure}
        
\caption{Evolution of a top-hat sample in SHPB after impact with a 4.8~m/s striker velocity. The average global strain rate is $1.2\times10^3$~s$^{-1}$. The overlay shows horizontal displacement.}
\label{fig:highrate}       % Give a unique label
\end{figure}

As before flat faced top-hat shaped sample of IMI834 with dimensions given in Figure \ref{fig:tophat} was used. A striker velocity of 4.8$\pm0.1$~m/s, measured optically, was selected to produce an average global strain rate of $1.2\times10^3$~s$^{-1}$. The 3-D DIC system was used to measure the surface deformation, and in particular the shear strain. Figure \ref{fig:highrate}a-c shows the sample with an overlay of the calculated x-displacement (that is displacement in the compressive direction) while Figure \ref{fig:highrate}d shows a spatial average of the shear strain across the high shear region. As the global compressive strain is increased a region of high shear can be seen to develop. In Figure \ref{fig:highrate_last} the deformation of the target causes the speckle pattern to become distorted to such a degree that the image correlation technique is no longer able to identify the correct regions causing spatial anomalies in the displacement map. The width of the shear band, approximately $250~{\mu}$m, is less than observed in the low strain-rate tests. This is expected as the shear band characteristic width is a combination of thermal diffusivity and loading time \cite{bdodd,zhen}, in the high rate test heat is localised close to the region of maximum shear. As before, towards the end of the test the shear strain increases further to a maximum observed value of 0.16 and ultimately leads to sample failure. The width of the shear bands found in this work are larger than those found previously in Ti alloy samples \cite{bandwidth1, bandwidth2} and those predicted by crystal plasticity modelling \cite{zhen} which are typically in the range 10-50 $\mu$m, however, the top-hat setup is sensitive to geometric effects, in particular, the overlap of the leg with the body of the top-hat \cite{tophat1}. Future work will investigate how the shear-band width varies with parameters such as strain-rate, geometry and microstructure.  

\section{Conclusions}
\label{sec:conc}

Experimental mechanics is constantly investigating the performance of new materials under load as attempts are made to improve cost, efficiency and performance. A crucial metric is obtaining the stress-strain data from such tests, and the far field measurements typically obtained are supplemented with surface strain measurements. This data is then fed into constitutive material models to aid future development, and can often involve performing tests on non-uniform samples or unique sample geometries. A full 3-D DIC displacement map allows a complete and direct comparison with finite element analysis allowing greater confidence in obtaining elastoplastic parameters such as strength or strain hardening exponents \cite{3ddicfea}. 

A method for obtaining the full-field displacement of both in and out-of-plane motion from a single high frame-rate camera is demonstrated. Using a standard digital SLR (Canon EOS) and a high frame rate camera (Phantom V7.3) together with Infinity K2 DistaMax long-distance microscope lens enabled high resolution images to be taken. 

To achieve this we utilised two open-source codes. The Ncorr two dimensional DIC code is used to perform the correlations between a pair of 2-D images while the MATLAB camera calibration or CalTech camera calibration toolbox is used to calibrate the system through the chequerboard method. The single camera system provides improved temporal synchronisation and spatial stability while the open-source software provides flexibility and transparency without resorting to expensive commercial software or necessitating multiple high speed cameras. 

Tests were performed using a high resolution Canon EOS with 5184x3456 pixels as well as with the Phantom high speed camera at full (800x600) and reduced (240x184) pixel number. In each case the obtainable maximum accuracy in the DIC displacement tests was found to be around 200-300 times better than the optical resolution of the set-up, in line with previous work \cite{singlecamera1}. The software was then compared to the DaVis commercially available code using the same image sets. This removed any ambiguity around illumination, speckle quality, camera resolution or lens quality. The code reproduced the displacement to within 200 um and the maximum normal surface strain to within 0.2. 

The performance of the system was then demonstrated on two quasi-static strain tests. In the first test a millimeter  sized top-hat sample was compressed to induce large shear strains in the region between the legs and body of the top-hat sample. The high resolution of the DIC system enabled a 400~$\mu$m band of high shear strain to be identified and quantified before failure. The second quasi-static test utilised a dog-bone sample in a tension test to observe necking in the Z direction. Large displacements were observed and the necking was measured to within 50~$\mu$m uncertainty.  Finally, a high-strain test on the same top-hat samples as compressed quasi-statically was carried out on a split-Hopkinson pressure bar at a global strain rate of $1.2\times10^3$~s$^{-1}$. Images captured at 120 000 Hz showed a thinner region of high-shear approximately 250~$\mu$m wide, suggesting adiabatic conditions were reach in the high rate tests.

Future work will utilise the single camera system and codes to further investigate the physics surrounding dynamic events on the sub-millimeter scale such as grains, shear bands and localised surface strains. 

\section{ACKNOWLEDGEMENTS}

This research was supported by EPSRC grant Heterogeneous Mechanics in Hexagonal Alloys across Length and Time Scales (EP/K034332/1). In addition, the authors are very grateful to David Rugg and Rolls-Royce plc for provision of Ti material.

\end{document}